\documentclass[10pt,a4paper,twoside,bibliography=totocnumbered]{article}

\usepackage{xcolor}

\usepackage{graphicx}

\usepackage{amsmath}
\usepackage{amssymb}

\usepackage{geometry}

\usepackage{natbib}
\renewcommand{\vec}{\boldsymbol}

\begin{document}

% title page
\title{{\huge \textcolor{black}{Errors-In-Variables Model Fitting for Partially Unpaired Data Utilizing Mixture Models }\\}}

\author{Wolfgang Hoegele$^{1}$ and Sarah Brockhaus$^{1}$\vspace*{0.3cm}\\
{\normalsize $^{1}$ Munich University of Applied Sciences HM}\\{\normalsize Department of Computer Science and Mathematics}\\{\normalsize Lothstraße 64, 80335 München, Germany}\vspace*{0.2cm}\\
{\normalsize \texttt{wolfgang.hoegele@hm.edu}}\vspace*{0.2cm}}

\date{\today}

\maketitle
\thispagestyle{empty}

\section*{Abstract}
We introduce a general framework for regression in the errors-in-variables regime, allowing for full flexibility about the dimensionality of the data, observational error probability density types, the (nonlinear) model type and the avoidance of ad-hoc definitions of loss functions. In this framework, we introduce model fitting for partially unpaired data, i.e. for given data groups the pairing information of input and output is lost (semi-supervised). This is achieved by constructing mixture model densities, which directly model the loss of pairing information allowing inference.
In a numerical simulation study linear and nonlinear model fits are illustrated as well as a real data study is presented based on life expectancy data from the world bank utilizing a multiple linear regression model.
These results show that high quality model fitting is possible with partially unpaired data, which opens the possibility for new applications with unfortunate or deliberate loss of pairing information in data. \medskip

\textbf{Keywords:} Errors-In-Variables; Mixture Models; Model Fitting; Semi-Supervised; Total Least Squares\medskip

\section*{About the Authors}

Dr. Högele is a Full Professor of Applied Mathematics and Computational Science at the Department of Computer Science and Mathematics at the Munich University of Applied Sciences HM, Germany. \smallskip

Dr. Brockhaus is a Full Professor of Applied Mathematics and Statistics at the Department of Computer Science and Mathematics at the Munich University of Applied Sciences HM, Germany.\bigskip\\

\textit{This manuscript is accepted and will be published in STATISTICS (Taylor \& Francis), 2024}\medskip

\textit{This manuscript appears also on ArXiv.org\\
arxiv:2406.18154 [stat.ME], https://doi.org/10.48550/arXiv.2406.18154}

\thispagestyle{empty}

% table of contents
\newpage
\thispagestyle{empty}
\tableofcontents

\newpage
\section{Introduction}

Parametric model fitting is a standard task in many applications starting from problem specific models with a few meaningful parameters to huge, flexible models (such as in the training phase of ANNs) \citep{zhang_parameter_1997,bishop_pattern_2006}. The general idea is that a defined parametric model family is fitted to given input / output data (i.e. supervised learning) with the purpose to estimate the \textit{best fit} parameters of the model representing the data. Typically, loss functions are defined for this purpose, such as the squared or absolute losses \citep{wang_comprehensive_2022}. Practical difficulties of model fitting are i) finding the appropriate model family and their \textit{parameters}, ii) defining an adequate loss function, iii) applying an efficient optimization algorithm and iv) dealing with data deficiencies (such as outliers, strong noise, incompleteness, partially lost pairing information, etc.).  

A possibility to avoid the definition of an \textit{ad-hoc} loss function is the modeling of  known uncertainties in the data and applying Maximum Likelihood (ML) approaches, which, in consequence, lead inherently to data driven loss functions. For example, data uncertainties can appear only in the output data (such as for ordinary least squares) or in input and output data, also known as the \textit{errors-in-variables} approaches. A well-known connection is between a normal distributed error in the output and the squared loss function. These approaches can always be extended to Bayesian estimations if prior distributions are assumed for the model parameters leading to Maximum A Posteriori (MAP) or Minimum Mean Squared Error (MMSE) algorithms based on the posterior distribution \citep{hoegele_bayesian_2013}.

\begin{figure}[h!]
\centering
\includegraphics[width=13cm]{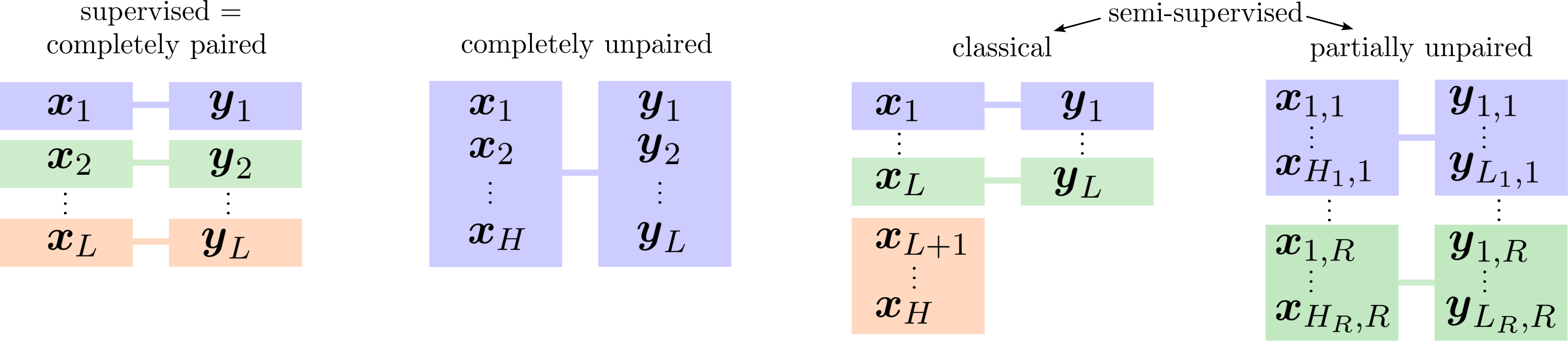}
\caption{Nomenclature of data configurations: \textit{completely paired}, \textit{completely unpaired} and \textit{semi-supervised} with the \textit{classical} and the \textit{partially unpaired} configuration.  }
\label{fig:DataExplanation}
\end{figure}

\textit{Unpaired / unlabeled} data (also \textit{broken sampling} in regression) can occur in many applications and is focus of current research, e.g. \citep{bai_broken_2005,liang_use_2007,wang_paired_2022}. In this work, we understand by \textit{partially unpaired} data (under the term \textit{semi-supervised data}) specifically that for parts of the data the one-to-one pairing of input and output data is missing, but we still have paired subgroups in the data. \textit{Semi-supervised} data often is understood as having additional input data without output data, e.g. \citep{kostopoulos_semi-supervised_2018,qi_small_2022}. This is different to the partially unpaired data in this paper and we regard this as a different flavor of \textit{semi-supervised} data since we lose significant information compared to a fully \textit{supervised} framework, but we still have (possibly weak) input/output relations in contrast to \textit{unsupervised learning}. See Figure \ref{fig:DataExplanation} for an illustration of different pairing information between input $\vec{x}$ and output $\vec{y}$. It is demonstrated that mixtures of labeled and unlabeled data can improve the predictive performance in regression problems \citep{liang_use_2007}. There are different strategies to deal with such deficiencies, such as altering the data set by data imputation approaches, e.g. see \citep{bennett_how_2001,sterne_multiple_2009}, or focusing on unordered subsets in hypothesis testing \citep{wang_paired_2022}. We regard the work of Liang \textit{et al.} \citep{liang_use_2007} considering a general predictive Bayesian frameworks for mixed labeled and unlabeled data as closest to our goal. Although the work provides a very general framework for regression and classification, it misses the configurational complexity of partially unpaired data. In consequence, one task of this paper is to incorporate the largest possible variety of missing pairing information transparently to model fitting by avoiding data alteration (deletion or imputation) and actively constructing mixture model probability densities accurately representing the partial pairing.

Mixture Models are probability density functions, which are composed of the weighted sum of elementary probability densities. They are applied in a variety of applications, especially Gaussian Mixture Models (GMM) are very popular, which are the weighted sum of Gaussian densities, e.g. \citep{liang_use_2007}. A main task is in the literature to find the GMM description of a given data set utilizing expectation maximization (EM) algorithms, also for the task of missing data deficiencies \citep{michael_mixture_2020,mccaw_fitting_2022}.
To the knowledge of the authors, a general framework for constructing mixture models for dealing with the high complexity of corrupt / incomplete pairing information of the data with the goal to support a model fitting problem is missing in the literature. Closely related approaches of utilizing mixture models in applied mathematics with lost pairing information are presented for computer vision in order to deal with unknown feature correspondences \citep{hoegele_stochastic-geometrical_2024-1} and for random equations with high combinatorial possibilities for the stochastic parameters \citep{hoegele_investigating_2024}.

\begin{figure}[h!]
\centering
\includegraphics[width=13cm]{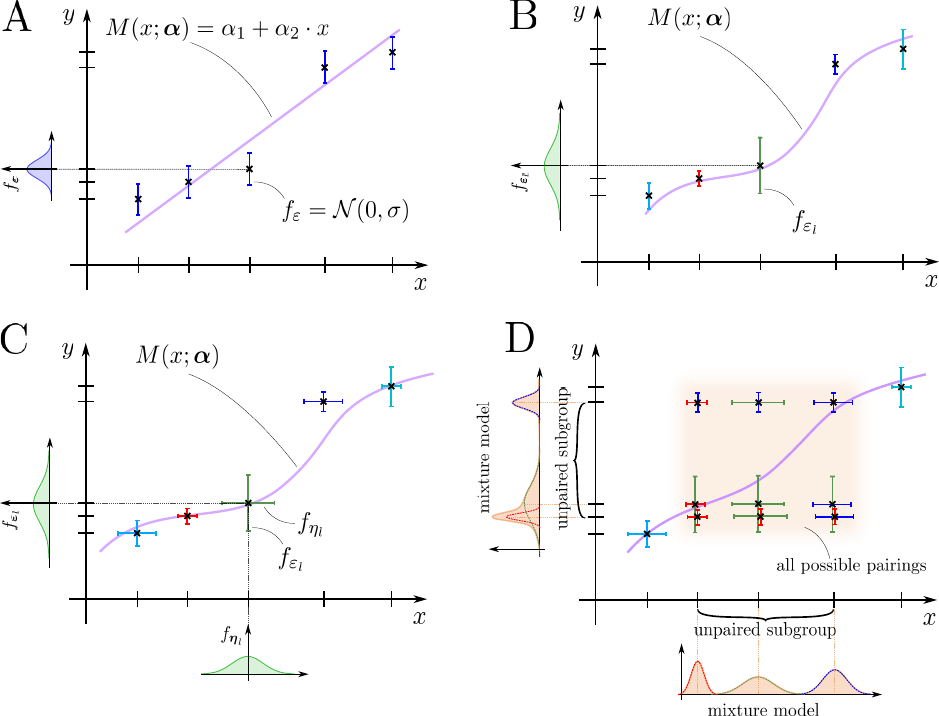}
\caption{Schematic illustration of the model fitting concepts in this paper for an input $x\in\mathbb{R}$ and output $y\in\mathbb{R}$, presented as steps of generalizations. A) Well--known \textit{ordinary least squares for a line fit} model, which corresponds to \textit{Maximum Likelihood} estimation with a constant Gaussian observational error density $f_{\boldsymbol{\varepsilon}}$ in $y$. B) Generalization of case (A) to a \textit{general nonlinear model} $M(x;\vec{\alpha})$ and \textit{general observational error density} $f_{\boldsymbol{\varepsilon}_l}$ in $y$ individual for each data point $l=1,..,L$. This is presented in \textit{Supplement \ref{sec:Deriv_errors_in_y_only}}. C) Generalization of (B) to the \textit{errors-in-variables} approach including observational error density functions $f_{\boldsymbol{\eta}_l}$ in $x$. This is presented in \textit{Section \ref{sec:ordered}}. D) Generalization of (C) to also include partially unpaired data subgroups, which contains all possible pairings of data within a subgroup (in this case $3\cdot 3 = 9$ pairings). This is realized with mixture models, which are the weighted sum of the data point probability densities. This is presented in \textit{Section \ref{sec:partiallyunordered}}.}
\label{fig:ConceptExplanation}
\end{figure}

In this paper, we propose to consider the problem of model fitting in a new conclusive way. A general probabilistic framework for fitting models in data based purely on \textit{observational error probability density functions} including \textit{errors-in-variables} is presented, which has direct relations to well-known standard methods for completely paired data, such as ordinary least squares, Deming regression \citep{deming_statistical_1964}, total least squares\citep{markovsky_overview_2007}, interval data regression as well as multiple linear regression. Observational error densities can have different reasons, for example, errors only in the output can be classical Gaussian measurement errors, but if regression is performed additionally with measured input data then both, input and output errors (\textit{errors-in-variables}), are typically described by Gaussians. Another example is \textit{interval data}, which can be represented by uniform densitiy functions and which can occur, e.g., in survey data, in particular, when asking for sensitive information like income. This framework will be generalized from \textit{supervised} to \textit{semi-supervised model fitting} by including (partially) unpaired data in one common line of stochastic argumentation. It is a key point that the presented derivations allow for full flexibility about i) the number and dimensions of the input / output data, ii) the type of individual error characteristics of each data point with \textit{errors-in-variables} utilizing general density functions, iii) the type of (linear or nonlinear) models which should be fitted and iv) the pairing information level. 
In the schematic Figure \ref{fig:ConceptExplanation} the concepts of this work are explained in an overview starting from the well--known ordinary least squares application of a line fit in \textit{Subfigure A} to the most general concept of this paper in \textit{Subfigure D} with a nonlinear fit in partially unpaired data. This presentation utilizes a one-dimensional input and output for illustrative purposes, which is generalized in the paper to arbitrary dimensions. 

In summary, there are two main goals of this paper: I) presenting a general stochastic argumentation framework for model fitting (without \textit{ad-hoc} loss functions) (see Figure \ref{fig:ConceptExplanation}A--C) and II) presenting an extension of model fitting within this framework to partially unpaired data utilizing mixture models (see Figure \ref{fig:ConceptExplanation}D).  

In I) the main ideas of the stochastic argumentation are: a) Formulating the fitting problem as a Maximum Likelihood (ML) problem of difference random variables. b) Applying the law of total probability for densities wherever necessary to make sure that correct stochastic dependencies are utilized and identifying the density functions of the basic random variables in the ML problem. c) Presentation of the optimization problems by the resulting objective functions.
Points a) to c) are presented repeatedly for different fitting scenarios and their generality is a \textit{first main result} of this paper.

In II) the extension of model fitting to partially unpaired data is structured in the following three-step-approach: Presenting the cases for
\begin{itemize}
\item completely paired data sets as the standard case in model fitting (i.e., \textit{supervised learning}) (Section \ref{sec:ordered}).
\item completely unpaired data sets are introduced mathematically utilizing mixture model random variables to model fitting (Section \ref{sec:unordered}). In this extreme case it is impossible to model the relation between input and output.
\item partially unpaired data sets, which lie between the two previous extremes and include different levels of pairing (i.e., \textit{semi-supervised learning}) (Section \ref{sec:partiallyunordered}).
\end{itemize}
This structure is chosen in order to allow clear and separated lines of argumentation which eventually conclude in the \textit{second main result} of the paper.

In the results section, we demonstrate the applicability of this framework by simulation studies with Gaussian and uniform mixture models as observational error densities for a line fit (Section \ref{sec:res_line}) and a fit of anisotropic noisy data with a cubic polynomial (Section \ref{sec:res_nonlinear}). Further, in Section \ref{sec:RealDataStudy}, we will demonstrate how this argumentation can be applied to multiple linear regression for \textit{life expectancy} data from the world bank. These results demonstrate the importance of modeling the inherent uncertainties and the use of different levels of pairing information in data.

\section{Methods}

\subsection{General Nomenclature}

Throughout the paper, we utilize the following nomenclature:
\begin{itemize}
\item Observations are presented by input data $\vec{x}_l\in\mathbb{R}^k$ and output data $\vec{y}_l\in\mathbb{R}^m$ for $l=1,..,L$ as independent observations. 
\item We call the data set \textit{completely paired} if for every $l=1,..,L$ there is a unique correspondence between $\vec{x}_l$ and $\vec{y}_l$, typically written as tuples $(\vec{x}_l,\vec{y}_l)$.

We call the data \textit{completely unpaired} if there is no pairing at all, i.e. there is a set of $\vec{x}_h$ for $h=1,..,H$ and independently a set of $\vec{y}_l$ for $l=1,..,L$ and there is no information which $\vec{x}_h$ corresponds to which $\vec{y}_l$.

In consequence, \textit{partially unpaired} data are a mix of both extremes, i.e. we have $R$ subgroups of the data $\vec{x}_h$ and $\vec{y}_l$ and inside each subgroup there is no pairing information of the data (no correspondences) but it is guaranteed that no $\vec{x}_h$, or $\vec{y}_l$ respectively, of one subgroup corresponds to a $\vec{x}_h$, or $\vec{y}_l$ respectively, of another subgroup. This means, we have pairing information only on the level of subgroups. These data configurations are illustrated in Figure \ref{fig:DataExplanation}.

\item Probability density functions are denoted by $\vec{X} \sim f_{\vec{X}}(x)$, which are Lebesgue integrable $f_{\vec{X}}(x)\in L^1$ and contain standard cases such as the normal or uniform distribution. Dirac distributions are used for theoretical discussions to show the connection between undisturbed and disturbed observations.   

\item We denote explicit models to be fitted from $\vec{x}$ to $\vec{y}$ by functions $\vec{M}(\;\cdot\;;\vec{\alpha}) : \mathbb{R}^k\mapsto \mathbb{R}^m$ depending on the model parameters $\vec{\alpha}\in\mathbb{R}^N$. 
A toy example for an explicit model is the one-dimensional affine model of linear regression $\mathbb{R}\mapsto \mathbb{R}$ ($k=1,m=1,N=2$): $M(x;\vec{\alpha})=\alpha_1+\alpha_2\cdot x$. 

\item Following a flexible Bayesian view on random variables is essential in this work. In the classical perspective we have an undisturbed variable $\vec{y}^{\,\ast}$ (the true value), which is disturbed by an error random variable $\vec{\varepsilon}$ leading to the observation $\vec{y}$, in short: $\vec{y} := \vec{y}^{\,\ast} + \vec{\varepsilon}$. 
We interpret $\vec{y}$ as a new random variable of observations which has a shifted density function of $\vec{\varepsilon}$ (we interpret $\vec{y}^{\,\ast}$ as a random variable with Dirac distribution at the true value). In this work, we consequently take an alternative point of view and introduce the definition by the reformulation: $\vec{y}^{\,\ast} := \vec{y} - \vec{\varepsilon}$. 
This time we interpret $\vec{y}^{\,\ast}$ as a new random variable of the true values (as typical in Bayesian frameworks) which has a shifted density function of $\vec{\varepsilon}$ (this time, we interpret $\vec{y}$ as the random variable with Dirac distribution at the observed value). The meaning of $\vec{\varepsilon}$ in these two perspectives is different but related, capturing the uncertainty of observation with a different center. We consequently utilize the latter notation in the rest of the paper.

\item In the stochastic argumentation, we utilize the notation
\begin{align*}
 f_{ \bigcap\limits_{l=1}^L \vec{Z}_l }(\vec{z}) :=  f_{ \vec{Z}_1,\dots,\vec{Z}_L }(\vec{z})
\end{align*}
with the meaning of the \textit{common density function} of all individual random variables $\vec{Z}_1,\dots,\vec{Z}_L$. 

\item In the following derivations, we focus on the \textit{argmax}/\textit{argmin} of an expression. Since the \textit{argmax}/\textit{argmin} is independent of the application of strictly monotonic increasing functions, we neglect those in the course of argumentation, i.e. for $c\in\mathbb{R}^+$ we will write
\begin{align*}
\text{argmax}_{\;\vec{\alpha}} \;\; c\cdot f(\vec{\alpha}) = \text{argmax}_{\;\vec{\alpha}} \; f(\vec{\alpha}) = \text{argmax}_{\;\vec{\alpha}} \; \ln\left(f(\vec{\alpha})\right) \;.
\end{align*}

\end{itemize}

\subsection{Model Fit with Completely Paired Data}
\label{sec:ordered}

The observations in this section are of type $(\vec{x}_l,\vec{y}_l)$ as tuples for $l=1,..,L$. Standard approaches, such as ordinary least squares (errors in $\vec{y}$ only) in this argumentation framework are presented in \textit{Supplement \ref{sec:Deriv_errors_in_y_only}}.

We introduce disturbances in the input and output data with the notation (which is referred in literature to \textit{errors-in-variables} \citep{markovsky_overview_2007}):
\begin{align}
\vec{x}_l^{\,\ast} &:= \vec{x}_l - \vec{\eta}_l\label{equ:EIV_x_ordered}\\
\vec{y}_l^{\,\ast} &:= \vec{y}_l - \vec{\varepsilon}_l\label{equ:EIV_y_ordered}
\end{align}
with $\vec{x}_l^{\,\ast}\in\mathbb{R}^k$ and $\vec{y}_l^{\,\ast}\in\mathbb{R}^m$ the random variables of true values, and the uncertainty random variables $\vec{\eta}_l \sim f_{\vec{\eta}_l}(\vec{s}) : \mathbb{R}^k\mapsto \mathbb{R}$ and $\vec{\varepsilon}_l \sim f_{\vec{\varepsilon}_l}(\vec{s}) : \mathbb{R}^m\mapsto \mathbb{R}$ independent for all $l=1,..,L$. We are interested in the case where the model is correctly chosen so that the true $\vec{x}_l^{\,\ast}$ predicts the true $\vec{y}_l^{\,\ast}$:
\begin{align*}
\vec{M}(\vec{x}_l^{\,\ast};\vec{\alpha}) &\stackrel{d}{=} \vec{y}_l^{\,\ast} \quad\forall\,l=1,..,L\\
\vec{M}(\vec{x}_l-\vec{\eta}_l;\vec{\alpha}) &\stackrel{ d}{=} \vec{y}_l - \vec{\varepsilon}_l \quad\forall\,l=1,..,L\;,
\end{align*}
with $\stackrel{d}{=}$ meaning equality \textit{in distribution}. Due to $\vec{\eta}_l$ and $\vec{\varepsilon}_l$ being random variables, the differences of right and left side $\vec{M}(\vec{x}_l-\vec{\eta}_l;\vec{\alpha}) - \vec{y}_l + \vec{\varepsilon}_l$ are interpreted as difference random variables for all $l=1,..,L$ whose density function values should have the highest possible value at $\vec{0}\in\mathbb{R}^m$ to achieve the most probable equality leading to the Maximum Likelihood approach:
\begin{align}
\Rightarrow \;&\text{argmax}_{\;\vec{\alpha}}\; f_{ \bigcap\limits_{l=1}^L \left[\, \vec{M}(\vec{x}_l-\vec{\eta}_l;\vec{\alpha}) - \vec{y}_l + \vec{\varepsilon}_l \,\right]}(\vec{0})\nonumber\\
\;{=}\;& \text{argmax}_{\;\vec{\alpha}}\; \prod\limits_{l=1}^L f_{\vec{M}(\vec{x}_l-\vec{\eta}_l;\vec{\alpha}) - \vec{y}_l + \vec{\varepsilon}_l }(\vec{0})\nonumber\qquad\text{\small (independency of $\vec{\eta}_l$, $\vec{\varepsilon}_l$ $\forall\,l$)} \\
\;{=}\;& \text{argmax}_{\;\vec{\alpha}}\; \prod\limits_{l=1}^L \;\int\limits_{\mathbb{R}^k} f_{\vec{M}(\vec{x}_l-\vec{s};\vec{\alpha}) - \vec{y}_l + \vec{\varepsilon}_l }(\vec{0})\cdot f_{\vec{\eta}_l}(\vec{s})\;\text{d}\vec{s}\nonumber\qquad\text{\small (law of total probility)}\\
\;{=}\;& \text{argmax}_{\;\vec{\alpha}}\; \prod\limits_{l=1}^L \;\int\limits_{\mathbb{R}^k} f_{\vec{\varepsilon}_l }(\vec{y}_l-\vec{M}(\vec{x}_l-\vec{s};\vec{\alpha}))\cdot f_{\vec{\eta}_l}(\vec{s})\;\text{d}\vec{s}\;\label{equ:ordered_xy2}\qquad\text{\small (shifted $\vec{\varepsilon}_l$)}\\
\;{=}\;& \text{argmax}_{\;\vec{\alpha}}\; \prod\limits_{l=1}^L \;\int\limits_{\mathbb{R}^k} f_{\vec{\varepsilon}_l }(\vec{y}_l-\vec{M}(\vec{s};\vec{\alpha}))\cdot f_{\vec{\eta}_l}(\vec{x}_l-\vec{s})\;\text{d}\vec{s}\;\label{equ:ordered_xy}\qquad\text{\small (integral shift)}
\end{align}
Applying the \textit{law of total probability} allows recovering the observation density functions in the final ML expression. In the following, we present examples of this general formula (\ref{equ:ordered_xy}) (or equivalently Equation (\ref{equ:ordered_xy2}) if beneficial).\smallskip

\textit{Remark:} By setting $f_{\vec{\eta}_l}(\vec{s}) = \delta(\vec{s})$ (the Dirac distribution), we allow no variation of the $\vec{x}_l^{\,\ast}$-values and, in consequence, get the equation of ordinary least squares (\textit{Supplement \ref{sec:Deriv_errors_in_y_only}}) by applying the sifting property. In consequence, this can be regarded as a true generalization of errors in $\vec{y}$ only.

\subsubsection{Example: Fitting a Line and Gaussian Disturbance}

In the line of total least squares \citep{markovsky_overview_2007}, we are introducing Gaussian disturbances by ${\eta}_l \sim \mathcal{N}(0,\,\sigma_{{\eta}}^{2})(s)\;\forall l=1,..,L$ and ${\varepsilon}_l \sim \mathcal{N}(0,\,\sigma_{{\varepsilon}}^{2})(s)\;\forall l=1,..,L$. Utilizing the one-dimensional affine model $M(x;\vec{\alpha})=\alpha_1+\alpha_2\cdot x$ and inserting it in Equation (\ref{equ:ordered_xy}), we get
\begin{align}
\Rightarrow \quad& \text{argmax}_{\;\vec{\alpha}}\quad \prod\limits_{l=1}^L \;\int\limits_{\mathbb{R}} \text{e}^{-\frac{1}{2\,\sigma_{{\varepsilon}}^2}(y_l-\alpha_1-\alpha_2\cdot s)^2 -\frac{1}{2\,\sigma_{{\eta}}^2} (x_l-s)^2 }\;\text{d}s\label{equ:deming_alternative_01}\\
= \quad& \text{argmax}_{\;\vec{\alpha}}\quad \;\prod\limits_{l=1}^L \;  \frac{\text{e}^{-\frac{(\alpha_1+\alpha_2\cdot x_l - y_l)^2}{2\,(\alpha_2^2\,\sigma_{{\eta}}^2 + \sigma_{{\varepsilon}}^2)}}}{\sqrt{\alpha_2^2\,\sigma_{{\eta}}^2 + \sigma_{{\varepsilon}}^2}}\nonumber\\
%= \quad& \text{argmax}_{\;\vec{\alpha}}\quad \text{e}^{-\sum\limits_{l=1}^L\; \frac{(\alpha_1+\alpha_2\cdot x_l - y_l)^2}{2\,(\alpha_2^2\,\sigma_{{\eta}}^2 + \sigma_{{\varepsilon}}^2)}}\\
= \quad& \text{argmin}_{\;\vec{\alpha}}\quad  \frac{L}{2}\,\ln\left( \alpha_2^2\,\sigma_{{\eta}}^2 + \sigma_{{\varepsilon}}^2\right) + \sum\limits_{l=1}^L\; \frac{(\alpha_1+\alpha_2\cdot x_l - y_l)^2}{2\,(\alpha_2^2\,\sigma_{{\eta}}^2 + \sigma_{{\varepsilon}}^2)}\;.\label{equ:deming_alternative}
\end{align}
Since this is a Deming regression type of problem, the solution to this minimization is also closely related to the classical Deming regression. The similarities and differences are presented in \textit{Supplement \ref{sec:Examples_errors_in_x_and_y}}. 

\subsubsection{Example: Fitting a Hyperplane and Gaussian Disturbance (Errors-In-Variables Multiple Linear Regression)}

Further extending this argumentation to hyperplanes for $\vec{x}_l\in\mathbb{R}^k$ and ${y}_l\in\mathbb{R}$, and ${\vec{\eta}}_l \sim \mathcal{N}(\vec{0},\,\text{diag}(\sigma_{{\eta},1}^{2},..,\sigma_{{\eta},k}^{2}))(\vec{s})$ $\forall l=1,..,L$ and ${\varepsilon}_l \sim \mathcal{N}(0,\,\sigma_{{\varepsilon}}^{2})(s)\;\forall l=1,..,L$ for fitting an affine hyperplane model $M(\vec{x};\vec{\alpha})=\alpha_1+\sum\limits_{n=1}^{k} \alpha_{n+1}\cdot x_n$ leads (utilizing Equation (\ref{equ:ordered_xy})) to the general optimization problem
\begin{align*}
\Rightarrow \quad&\text{argmax}_{\;\vec{\alpha}}\quad \prod\limits_{l=1}^L \;\int\limits_{\mathbb{R}^k} \text{e}^{ -\frac{1}{2\,\sigma_{\varepsilon}^2} \left( {y}_l-\alpha_1-\sum\limits_{n=1}^{k}\alpha_{n+1}\,s_n \right)^2  - \frac{1}{2} \left( \sum\limits_{n=1}^{k} \frac{(x_{l,n}-s_n)^2}{\sigma_{\eta,n}^2} \right)  }\;\text{d}\vec{s}\\
=\quad & \text{argmax}_{\;\vec{\alpha}}\quad  \prod\limits_{l=1}^L \;  \frac{\text{e}^{- \frac{\left(\alpha_1+\sum\limits_{n=1}^{k}\alpha_{n+1}\cdot x_{l,n} - y_l\right)^2}{2\,\left(\left(\sum\limits_{n=1}^k \alpha_{n+1}^2\,\sigma_{{\eta,n}}^2\right) + \sigma_{{\varepsilon}}^2\right)}}}{\sqrt{\left(\sum\limits_{n=1}^k \alpha_{n+1}^2\,\sigma_{{\eta,n}}^2\right) + \sigma_{{\varepsilon}}^2}}\\
=\quad & \text{argmin}_{\;\vec{\alpha}}\quad  \frac{L}{2}\,\ln\left( \left(\sum\limits_{n=1}^k \alpha_{n+1}^2\,\sigma_{{\eta,n}}^2\right) + \sigma_{{\varepsilon}}^2 \right) + \sum\limits_{l=1}^L \; \frac{\left(\alpha_1+\sum\limits_{n=1}^{k}\alpha_{n+1}\cdot x_{l,n} - y_l\right)^2}{2\,\left(\left(\sum\limits_{n=1}^k \alpha_{n+1}^2\sigma_{{\eta,n}}^2 \right)\,+ \sigma_{{\varepsilon}}^2\right)}\;.
\end{align*}

\subsubsection{Connection to Interval Data Regression}

Interval data is defined as data for which only the borders of an interval in which the true data point lies are observed. Performing model fitting for such data is an active field of research, e.g. for multilinear linear regression models \citep{lima_neto_centre_2008,souza_parametrized_2017}.

First, interval data for regression is defined the following way: For each data point coordinate $x_{l,i}$ ($i=1,..,k$) and $y_{l,j}$ ($j=1,..,m$), respectively, we only know the interval borders, i.e. $x_{l,i}\in[\underline{x_{l,i}},\overline{x_{l,i}}]$ and $y_{l,i}\in[\underline{y_{l,j}},\overline{y_{l,j}}]$, which are independently measured.
This information can be interpreted as a uniform distribution with probability mass inside the interval and zero outside. By introducing $\vec{x}_{l} := \frac{1}{2}\left( \underline{\vec{x}_{l}}+\overline{\vec{x}_{l}}\right)$ and $\vec{y}_{l} := \frac{1}{2}\left( \underline{\vec{y}_{l}}+\overline{\vec{y}_{l}}\right)$, and $v_{l,i} := \frac{1}{2}\left(\overline{x_{l,i}} - \underline{x_{l,i}}\right)$ and $w_{l,j} := \frac{1}{2}\left(\overline{y_{l,j}} - \underline{y_{l,j}}\right)$, this is equivalent to the general description of Equations (\ref{equ:EIV_x_ordered}) and (\ref{equ:EIV_y_ordered}) with
\begin{align*}
\vec{\eta}_{l} &\sim \prod\limits_{i=1}^k U_{\left[-v_{l,i},v_{l,i}\right]}(s_i)\\
\vec{\varepsilon}_{l} &\sim \prod\limits_{j=1}^m U_{\left[-w_{l,j},w_{l,j}\right]}(s_j)\;,
\end{align*}
with $U_{[a,b]}(s)$ the density function of the uniform distribution on the interval $[a,b]$.
This is obvious, since we can define an interval by either the two interval borders or the midpoint and its half width.

Second, this means we can apply Equation (\ref{equ:ordered_xy2}) for fitting a model $\vec{M}$ into that interval data, leading to
\begin{align*}
\Rightarrow\;\text{argmax}_{\;\vec{\alpha}}\quad \prod\limits_{l=1}^L \;\int\limits_{\mathbb{R}^k} &\prod\limits_{j=1}^m U_{\left[-w_{l,j},w_{l,j}\right]}\left(\frac{1}{2}\left( \underline{y_{l,j}}+\overline{{y}_{l,j}}\right)-M_j\left(\frac{1}{2}\left( \underline{\vec{x}_{l}}+\overline{\vec{x}_{l}}\right)-\vec{s};\vec{\alpha}\right)\right)\\
\cdot&  \prod\limits_{i=1}^k U_{\left[-v_{l,i},v_{l,i}\right]}(s_i)\;\text{d}\vec{s}\;.
\end{align*}
This can be further simplified to the $\text{argmax}_{\;\vec{\alpha}}$ of
\begin{align*}
\prod\limits_{l=1}^L \;\frac{1}{\prod\limits_{i=1}^k v_{l,i} }\cdot\int\limits_{\stackrel{\left[-v_{l,1},v_{l,1}\right]\times}{\dots\times \left[-v_{l,k},v_{l,k}\right]}} &\prod\limits_{j=1}^m U_{\left[-w_{l,j},w_{l,j}\right]}\left(\frac{1}{2}\left( \underline{y_{l,j}}+\overline{{y}_{l,j}}\right)-M_j\left(\frac{1}{2}\left( \underline{\vec{x}_{l}}+\overline{\vec{x}_{l}}\right)-\vec{s};\vec{\alpha}\right)\right)\;\text{d}\vec{s}\;.
\end{align*}
For each $\vec{s}$ the integrand is either zero or the positive normalization constant of the density of $\vec{\varepsilon}_l$, leading to a $k$-dimensional constant region for the integrand, whose volume is integrated over the $k$-dimensional box $\left[-v_{l,1},v_{l,1}\right]\times\dots\times \left[-v_{l,k},v_{l,k}\right]$. This means, the resulting optimization is searching for $\vec{\alpha}$ which maximizes the overlapping volume of the $k$-dimensional region with the $k$-dimensional box for all data points $l=1,..,L$ under consideration of the weights $v_{l,i}$ and $w_{l,i}$. This is an intuitive general understanding of model fitting with interval data.

\subsubsection{Example: Linear Interval Data Regression}

Utilizing the one-dimensional affine model $M(x;\vec{\alpha})=\alpha_1+\alpha_2\cdot x$, we can further derive
\begin{align*}
\Rightarrow &\;\text{argmax}_{\;\vec{\alpha}}\quad \prod\limits_{l=1}^L \;\frac{1}{v_l}\cdot\int\limits_{\left[-v_{l},v_{l}\right]}  U_{\left[-w_{l},w_{l}\right]}\left(\frac{1}{2}\left( \underline{y_{l}}+\overline{{y}_{l}}\right)-\left(\alpha_1+\alpha_2\,\left(\frac{1}{2}\left( \underline{x_{l}}+\overline{x_{l}}\right)-s\right)\right)\right)\;\text{d}s\\
\stackrel{\alpha_2\neq 0}{\Rightarrow} &\;\text{argmax}_{\;\vec{\alpha}}\quad \prod\limits_{l=1}^L \;\frac{1}{v_l}\cdot\int\limits_{\left[-v_{l},v_{l}\right]}  \frac{1}{2\,w_l}\,\chi_{\left[ c_{l,\text{min}}(\vec{\alpha}),c_{l,\text{max}}(\vec{\alpha})\right]}\left(s\right)\;\text{d}s
\end{align*}
with the abbreviations $c_{l,\pm}(\vec{\alpha}):=\frac{1}{2}\left( \underline{x_{l}}+\overline{x_{l}}\right)+\frac{1}{\alpha_2}\left(\alpha_1-\frac{1}{2}\left( \underline{y_{l}}+\overline{{y}_{l}}\right)\pm w_l\right)$, $c_{l,\text{min}} = \text{min}(c_{l,\pm})$, $c_{l,\text{max}} = \text{max}(c_{l,\pm})$ and $\chi_{[a,b]}(s)$ the \textit{characteristic function} ($1$ if $s\in[a,b]$, else $0$).
This can further be simplified to
\begin{align*}
\Rightarrow &\;\text{argmax}_{\;\vec{\alpha}}\quad \prod\limits_{l=1}^L \;\frac{1}{v_l\cdot w_l}\cdot\text{max}\left[\;\text{min}\left[ v_l, c_{l,\text{max}}(\vec{\alpha})  \right] - \text{max}\left[ -v_l,c_{l,\text{min}}(\vec{\alpha})  \right], 0\;\right]\;.
\end{align*}
For an example fit according to this formula, see the results Section \ref{sec:res_line}

\textit{Remark:} With this framework of argumentation, we can also introduce uncertainty about the knowledge of the integral borders in a transparent way by not assuming a strict uniform distribution, but a distribution blurred at the borders.

\subsection{Model Fit with Completely Unpaired Data}
\label{sec:unordered}

In this section, we are losing the property of tuples, i.e. observations of type $\vec{x}_h$ for $h=1,..,H$ and $\vec{y}_l$ for $l=1,..,L$ are unpaired. Only a set of $\vec{x}_h$ and a set of $\vec{y}_l$ observations are available. Please note, even the sizes $H$ and $L$ can be different.
Although there is theoretical research about the usability of such \textit{broken sampling} data sets, e.g. \citep{bai_broken_2005}, obviously such data will only lead to very limited regression results if there are no further assumptions about the involved probability densities since we only have marginal distributions. Nonetheless, we want to introduce a formulation by mixture models for this case, which will later be utilized for partially unpaired data directly. 
We start this argumentation with possible disturbances in input and output data: 
\begin{align*}
\vec{x}_h^{\,\ast}  &:= \vec{x}_h - \vec{\eta}_h\\
\vec{y}_l^{\,\ast}  &:= \vec{y}_l - \vec{\varepsilon}_l
\end{align*}
with $\vec{x}_h^{\,\ast}\in\mathbb{R}^k$ and $\vec{y}_l^{\,\ast}\in\mathbb{R}^m$ the random variables of the true values, and the uncertainty random variables $\vec{\eta}_h \sim f_{\vec{\eta}_h}(\vec{s}) : \mathbb{R}^k\mapsto \mathbb{R}$ and $\vec{\varepsilon}_l \sim f_{\vec{\varepsilon}_l}(\vec{s}) : \mathbb{R}^m\mapsto \mathbb{R}$ independent for all $h=1,..,H$ and $l=1,..,L$. A new step is now to introduce the two mixture model random variables
\begin{align*}
\vec{X}^{\ast} \sim f_{\vec{X}^{\ast}}(\vec{s}) &= \frac{1}{H} \sum\limits_{h=1}^H f_{\vec{x}_h-\vec{\eta}_h}(\vec{s}) = \frac{1}{H} \sum\limits_{h=1}^H f_{\vec{\eta}_h}( \vec{x}_h-\vec{s} )\\
\vec{Y}^{\ast} \sim f_{\vec{Y}^{\ast}}(\vec{s}) &= \frac{1}{L} \sum\limits_{l=1}^L f_{\vec{y}_l-\vec{\varepsilon}_l}(\vec{s}) = \frac{1}{L} \sum\limits_{l=1}^L f_{\vec{\varepsilon}_l}( \vec{y}_l-\vec{s} )\;,
\end{align*}
which exactly contain the ignorance of the pairing, i.e. all $\vec{x}_h$ and $\vec{y}_l$ observations are present in these mixture models at once. We follow the same technical argumentation as in the previous section:
\begin{align*}
\vec{M}(\vec{X}^{\ast};\vec{\alpha}) &\stackrel{ d}{=} \vec{Y}^{\ast}\;.
\end{align*}
Due to $\vec{X}^{\ast}$ and $\vec{Y}^{\ast}$ being random variables, the difference of right and left side $\vec{M}(\vec{X}^{\ast};\vec{\alpha}) - \vec{Y}^{\ast}$ is again interpreted as a difference random variable, whose density function value should have highest value at $\vec{0}\in\mathbb{R}^m$. This leads to the ML approach:
\begin{align}
\Rightarrow \;&\text{argmax}_{\;\vec{\alpha}}\; f_{ \vec{M}(\vec{X}^{\ast};\vec{\alpha}) - \vec{Y}^{\ast}}(\vec{0})\nonumber\\
\;{=}\;& \text{argmax}_{\;\vec{\alpha}}\; \int\limits_{\mathbb{R}^k} f_{\vec{M}(\vec{s};\vec{\alpha}) - \vec{Y}^{\ast} }(\vec{0})\cdot f_{\vec{X}^{\ast}}(\vec{s})\;\text{d}\vec{s}\nonumber\qquad\text{ \small (law of total probility)}\\
\;{=}\;& \text{argmax}_{\;\vec{\alpha}}\; \int\limits_{\mathbb{R}^k} f_{\vec{Y}^{\ast} }(\vec{M}(\vec{s};\vec{\alpha}))\cdot f_{\vec{X}^{\ast}}(\vec{s})\;\text{d}\vec{s}\label{equ:unordered_xy2}\qquad\text{\small (shifted $\vec{Y}^{\ast}$)}\\
\;{=}\;& \text{argmax}_{\;\vec{\alpha}}\; \int\limits_{\mathbb{R}^k} \left( \frac{1}{L}\sum\limits_{l=1}^L f_{\vec{\varepsilon}_l}( \vec{y}_l-\vec{M}(\vec{s};\vec{\alpha}) ) \right) \cdot \left( \frac{1}{H}\sum\limits_{h=1}^H f_{\vec{\eta}_h}( \vec{x}_h-\vec{s} )\right) \;\text{d}\vec{s}\nonumber\qquad \text{\small (def. of $\vec{X}^{\ast}$, $\vec{Y}^{\ast}$)}\\
{=}\;& \text{argmax}_{\;\vec{\alpha}}\; \frac{1}{L H}\sum\limits_{l=1}^L \sum\limits_{h=1}^H \;\int\limits_{\mathbb{R}^k} f_{\vec{\varepsilon}_l}( \vec{y}_l-\vec{M}(\vec{s};\vec{\alpha}) ) \cdot f_{\vec{\eta}_h}( \vec{x}_h-\vec{s} ) \;\text{d}\vec{s}\;.\label{equ:unordered_xy}
\end{align}
\textit{Remark:} The double sum in Equation (\ref{equ:unordered_xy}) takes care of all combinations of $\vec{x}_h$ and $\vec{y}_l$ coming directly from a strict stochastic derivation with these mixture model random variables. In the completely paired case with independent observations, a product appears in Equation (\ref{equ:ordered_xy}) which corresponds to this double sum for the completely unpaired case.

\textit{Remark:} This fit with completely unpaired data is practically useless. This means, there will be broad, probably non-distinct or multiple maxima in this objective function. Still, this argumentation helps in a theoretical perspective since it is applied directly to the partially unpaired data where we have a range of different levels of pairing information.

\subsection{Model Fit with Partially Unpaired Data}
\label{sec:partiallyunordered}
\label{sec:partunord}

This section introduces the argumentation for \textit{partially unpaired data}, which is a main result of this paper. For this case, we partition the $H$ observations $\vec{x}_h$ and $L$ observations $\vec{y}_l$ into $r=1,..,R$ disjoint independent groups, i.e. observations of group $r$ of type $\vec{x}_{h,r}$ for $h=1,..,H_r$ and $\vec{y}_{l,r}$ for $l=1,..,L_r$ are unpaired with $H_r$ and $L_r$ representing the number of elements in subgroup $r$. This means, we have a set of $\vec{x}_h$-values and a set of $\vec{y}_l$-values for each subgroup and we have pairing information on the group level. The number of input and output elements in each subgroup $H_r$ and $L_r$ are not necessarily the same. 
Again, we allow disturbances in $\vec{x}_h$ and $\vec{y}_l$: 
\begin{align*}
\vec{x}_h^{\,\ast} &:= \vec{x}_h - \vec{\eta}_h\\
\vec{y}_l^{\,\ast} &:= \vec{y}_l - \vec{\varepsilon}_l
\end{align*}
with $\vec{x}_h^{\,\ast}\in\mathbb{R}^k$ and $\vec{y}_l^{\,\ast}\in\mathbb{R}^m$ the random variables of the true values, and the independent uncertaintiy random variables $\vec{\eta}_h \sim f_{\vec{\eta}_h}(\vec{s}) : \mathbb{R}^k\mapsto \mathbb{R}$ and $\vec{\varepsilon}_l \sim f_{\vec{\varepsilon}_l}(\vec{s}) : \mathbb{R}^m\mapsto \mathbb{R}$ ($l=1,..,L$). The main argument for dealing with unpaired data is presented by mixture models, i.e. we define ($r=1,..,R$):
\begin{align*}
\vec{X}^{\ast}_r \sim f_{\vec{X}^{\ast}_r}(\vec{s}) &= \frac{1}{H_r} \sum\limits_{h=1}^{H_r} f_{\vec{\eta}_{h,r}}( \vec{x}_{h,r}-\vec{s} )\\
\vec{Y}^{\ast}_r \sim f_{\vec{Y}^{\ast}_r}(\vec{s}) &= \frac{1}{L_r} \sum\limits_{l=1}^{L_r} f_{\vec{\varepsilon}_{l,r}}( \vec{y}_{l,r}-\vec{s} )\;.
\end{align*}
Following our standard line of argumentation, we get
\begin{align}
\vec{M}(\vec{X}^{\ast}_r;\vec{\alpha}) &\stackrel{d}{=} \vec{Y}^{\ast}_r \quad\forall r=1,..,R \label{equ:matching_requirement}\;,
\end{align}
and again focus on the difference random variables $\vec{M}(\vec{X}^{\ast}_r;\vec{\alpha}) - \vec{Y}^{\ast}_r$ for all $r=1,..,R$ at $\vec{0}$. Following the ML approach, we arrive at
\begin{align}
\Rightarrow \;&\text{argmax}_{\;\vec{\alpha}}\; f_{ \bigcap\limits_{r=1}^R\;\left[\, \vec{M}(\vec{X}^{\ast}_r;\vec{\alpha}) - \vec{Y}^{\ast}_r\,\right]}(\vec{0})\nonumber\\
\;{=}\;& \text{argmax}_{\;\vec{\alpha}}\; \prod\limits_{r=1}^R \;f_{ \vec{M}(\vec{X}^{\ast}_r;\vec{\alpha}) - \vec{Y}^{\ast}_r}(\vec{0})\nonumber\qquad\text{\small (group independency)}\\
\;{=}\;& \text{argmax}_{\;\vec{\alpha}}\; \prod\limits_{r=1}^R \; \int\limits_{\mathbb{R}^k} f_{\vec{Y}^{\ast}_r }(\vec{M}(\vec{s};\vec{\alpha}))\cdot f_{\vec{X}^{\ast}_r}(\vec{s})\;\text{d}\vec{s}\label{equ:partiallyunordered_xy2}\qquad\text{\small (cp. equ. (\ref{equ:unordered_xy2}))}\\
\;{=}\;& {\; \text{argmax}_{\;\vec{\alpha}}\; \prod\limits_{r=1}^R \; \left[\;\frac{1}{L_r H_r}\sum\limits_{l=1}^{L_r} \sum\limits_{h=1}^{H_r} \;\int\limits_{\mathbb{R}^k} f_{\vec{\varepsilon}_{l,r}}( \vec{y}_{l,r}-\vec{M}(\vec{s};\vec{\alpha}) ) \cdot f_{\vec{\eta}_{h,r}}( \vec{x}_{h,r}-\vec{s} ) \;\text{d}\vec{s}\,\right]\;}\label{equ:partiallyunordered_xy}\qquad\text{{\small (cp. equ. (\ref{equ:unordered_xy}))}}
\end{align}
Equation (\ref{equ:partiallyunordered_xy}) is the most general formula we derive in this paper, since it contains the previous cases (completely paired data $R=H=L$ and completely unpaired data $R=1$). Most importantly, all other possibilities of partial pairing are contained in this equation. 
For example, ordinary least squares for paired data (a standard regression approach) is achieved by setting $R=L=H$ (equals group size $1$) and setting $f_{\vec{\eta}_h}(\vec{s})=\delta(\vec{s})$. 

\textit{Remark:} In the partially unpaired setup, we always work with input / output correspondences, only on a subgroup basis. A completely paired data subset (= \textit{supervised data}) is represented by subgroups of size one. An additional pure input data subset (= \textit{unsupervised}) can be approximated by neglecting the information about $\vec{Y}^{\ast}_r$, which corresponds to extremely flat $f_{\vec{\varepsilon}_{l,r}}$, arriving at the classical definition of \textit{semi-supervised}. This means, in Equation (\ref{equ:partiallyunordered_xy}) the first term in the integral $f_{\vec{\varepsilon}_{l,r}}$ gets essentially constant (independent of the prediction $\vec{M}(\vec{s};\vec{\alpha})$) and, therefore, this part becomes  practically noninformative with respect to the optimization on $\vec{\alpha}$. Thus, as expected, the completely unsupervised part of the data only on the input side can be neglected since it contains no information about the model parameters~$\vec{\alpha}$.

\textit{Remark:} An interesting point is that the same cannot be said about having unsupervised data on the output side, e.g. neglecting information about $\vec{X}^{\ast}_r$. This time $f_{\vec{\eta}_{h,r}}$ becomes a flat distribution, essentially leaving $f_{\vec{\varepsilon}_{l,r}}( \vec{y}_{l,r}-\vec{M}(\vec{s};\vec{\alpha}) )$ in the integral of Equation (\ref{equ:partiallyunordered_xy}) evaluated for all possible $\vec{s}$. This time the change of $\vec{\alpha}$ can have direct influence on the optimization, essentially taking care that the output values $\vec{y}_{l,r}$ are plausible / possible (i.e. in the probabilistically blurred image of $\vec{M}(\vec{s};\vec{\alpha})$) for a given parameter set $\vec{\alpha}$. This shows an asymmetry with respect the classical semi-supervised setup \citep{liang_use_2007}.

\subsubsection{Example: Fitting a Line and Gaussian Disturbance}

Gaussian disturbances in input and output variables with ${\eta}_h \sim \mathcal{N}(0,\,\sigma_{{\eta}}^{2})(s)\;\forall h=1,..,H$ and ${\varepsilon}_l \sim \mathcal{N}(0,\,\sigma_{{\varepsilon}}^{2})(s)\;\forall l=1,..,L$, and utilizing the one-dimensional affine model $M(x;\vec{\alpha})=\alpha_1+\alpha_2\cdot x$ and inserting it, we get by applying Equation (\ref{equ:partiallyunordered_xy})
\begin{align*}
\Rightarrow\quad & \text{argmax}_{\;\vec{\alpha}}\quad \prod\limits_{r=1}^R \; \left[\;\frac{1}{L_r H_r}\sum\limits_{l=1}^{L_r} \sum\limits_{h=1}^{H_r}\; \frac{\text{e}^{-\frac{(\alpha_1+\alpha_2\cdot x_{h,r} - y_{l,r})^2}{2\,(\alpha_2^2\,\sigma_{{\eta}}^2 + \sigma_{{\varepsilon}}^2)}}}{\sqrt{\alpha_2^2\,\sigma_{{\eta}}^2 + \sigma_{{\varepsilon}}^2}}\;\right]\;,
\end{align*}
which represents a solution to the Deming type problem for partially unpaired data. 

\subsubsection{Example: Fitting a Hyperplane and Gaussian Disturbance (Errors-In-Variables Multiple Linear Regression)}
\label{sec:ex:EIV_MultLinReg_PartUnord}

Gaussian disturbances in input and output variables with \\${\vec{\eta}}_h \sim \mathcal{N}(\vec{0},\,\text{diag}(\sigma_{{\eta},1}^{2},..,\sigma_{{\eta},k}^{2}))(\vec{s})\;\forall h=1,..,H$ and ${\varepsilon}_l \sim \mathcal{N}(0,\,\sigma_{{\varepsilon}}^{2})(s)\;\forall l=1,..,L$, and utilizing the $k$-dimensional affine model $M(\vec{x};\vec{\alpha})=\alpha_1+\sum\limits_{n=1}^{k} \alpha_{n+1}\cdot x_n$ and inserting it, we get by applying Equation (\ref{equ:partiallyunordered_xy})
\begin{align*}
\Rightarrow\quad & \text{argmax}_{\;\vec{\alpha}}\quad \prod\limits_{r=1}^R \; \left[\;\frac{1}{L_r H_r}\sum\limits_{l=1}^{L_r} \sum\limits_{h=1}^{H_r}\; \frac{\text{e}^{- \frac{\left(\alpha_1+\sum\limits_{n=1}^{k}\alpha_{n+1}\cdot x_{h,r,n} - y_{l,r}\right)^2}{2\,\left(\left(\sum\limits_{n=1}^k \alpha_{n+1}^2\,\sigma_{{\eta,n}}^2\right) + \sigma_{{\varepsilon}}^2\right)}}}{\sqrt{\left(\sum\limits_{n=1}^k \alpha_{n+1}^2\,\sigma_{{\eta,n}}^2\right) + \sigma_{{\varepsilon}}^2}}\;\right]\;,
\end{align*}
representing errors-in-variables multiple linear regression for partially unpaired data.

\subsubsection{Example: Linear Interval Data Regression}
As final example, we present interval data that are given with $\vec{x}_{h} := \frac{1}{2}\left( \underline{\vec{x}_{h}}+\overline{\vec{x}_{h}}\right)$ and $\vec{y}_{l} := \frac{1}{2}\left( \underline{\vec{y}_{l}}+\overline{\vec{y}_{l}}\right)$, and $v_{h,i} := \frac{1}{2}\left(\overline{x_{h,i}} - \underline{x_{h,i}}\right)$ and $w_{l,j} := \frac{1}{2}\left(\overline{y_{l,j}} - \underline{y_{l,j}}\right)$, and
\begin{align*}
\vec{\eta}_{h} &\sim \prod\limits_{i=1}^k U_{\left[-v_{h,i},v_{h,i}\right]}(s_i)\\
\vec{\varepsilon}_{l} &\sim \prod\limits_{j=1}^m U_{\left[-w_{l,j},w_{l,j}\right]}(s_j)\;.
\end{align*}
The one-dimensional affine model $M(x;\vec{\alpha})=\alpha_1+\alpha_2\cdot x$, leads to the $\text{argmax}_{\;\vec{\alpha}}$ of
\begin{align*}
\prod\limits_{r=1}^R \;\left[\;\frac{1}{L_r H_r}\sum\limits_{l=1}^{L_r} \sum\limits_{h=1}^{H_r} \;\frac{1}{v_{h,r}\cdot w_{l,r}}\cdot\text{max}\left[\; \text{min}\left[ v_{h,r}, c_{l,h,r,\text{max}}(\vec{\alpha})  \right] - \text{max}\left[ -v_{h,r},c_{l,h,r,\text{min}}(\vec{\alpha})  \right], 0\;\right]\;\right]\;,
\end{align*}
with $c_{l,h,r,\pm}(\vec{\alpha}):=\frac{1}{2}\left( \underline{x_{h,r}}+\overline{x_{h,r}}\right)+\frac{1}{\alpha_2}\left(\alpha_1-\frac{1}{2}\left( \underline{y_{l,r}}+\overline{{y}_{l,r}}\right)\pm w_{l,r}\right)$, $c_{l,h,r,\text{min}} = \text{min}(c_{l,h,r,\pm})$ and $c_{l,h,r,\text{max}} = \text{max}(c_{l,h,r,\pm})$.

\subsection{Extensions}

\subsubsection{Numerical Implementation of the General Formula for Partially Unpaired Data}
\label{sec:numericals}

We want to stress that the implementation of the general formula (\ref{equ:partiallyunordered_xy}) is not recommended if avoidable, due to typically high computational costs. A typical way to avoid this, is to work with specific probability density types or model families, such as presented in the examples following Equation (\ref{equ:partiallyunordered_xy}). 
For the general case, we provide the following implementation recommendations:

First, a beneficial numerical implementation strategy is to avoid the (possibly massive) multiplication in the general Equation (\ref{equ:partiallyunordered_xy}). In consequence, we rewrite this by applying the natural logarithm and multiplying it by $-1$ in order to generate a practically useful minimization problem
\begin{align*}
{\; \text{argmin}_{\;\vec{\alpha}}\quad -\sum\limits_{r=1}^R \; \ln\left( \frac{1}{L_r H_r}\left[\;\sum\limits_{l=1}^{L_r} \sum\limits_{h=1}^{H_r} \;\int\limits_{\mathbb{R}^k} f_{\vec{\varepsilon}_{l,r}}( \vec{y}_{l,r}-\vec{M}(\vec{s};\vec{\alpha}) ) \cdot f_{\vec{\eta}_{h,r}}( \vec{x}_{h,r}-\vec{s} ) \;\text{d}\vec{s}\,\right]\right)\;}\;.
\end{align*}
Second, although the formulation of Equation (\ref{equ:partiallyunordered_xy}) shows the combinatorics of the possible correspondences in unpaired data subsets, this is not an efficient way for implementation since it involves the approximation of $L_r\cdot H_r$ integrals for each subgroup $r$. It is recommended to utilize Equation (\ref{equ:partiallyunordered_xy2}), by first evaluating the mixture models $f_{\vec{Y}^{\ast}_r }$ and $f_{\vec{X}^{\ast}_r}$ for appropriate $\vec{s}$ for the numerical integration and then solving only one integral numerically for each subgroup, leading to the formula 
\begin{align*}
{\; \text{argmin}_{\;\vec{\alpha}}\quad -\sum\limits_{r=1}^R \; \ln\left(\;\;  \int\limits_{\mathbb{R}^k} f_{\vec{Y}^{\ast}_r }(\vec{M}(\vec{s};\vec{\alpha}))\cdot f_{\vec{X}^{\ast}_r}(\vec{s})\;\text{d}\vec{s} \; \right)\;}\;.
\end{align*}
Third, numerical approximation of the integral is necessary. For high dimensional input data dimensions $\vec{x}_h\in\mathbb{R}^k$ it is preferable to apply advanced Monte Carlo integration. For example, if we utilize $p=1,..,P$ samples $\vec{s}_{r,p}$ drawn from the mixture model density $f_{\vec{X}^{\ast}_r}$, then we can apply Monte Carlo integration with the formula
\begin{align}
{\; \text{argmin}_{\;\vec{\alpha}}\quad -\sum\limits_{r=1}^R \; \ln\left(\;\frac{1}{P}  \sum\limits_{p=1}^P f_{\vec{Y}^{\ast}_r }(\vec{M}(\vec{s}_{r,p};\vec{\alpha})) \; \right)\;}\;,
\label{equ:MonteCarlo}
\end{align}
which increases computational efficiency significantly.

Fourth, the choice of optimization algorithm depends strongly on the dimensionality of the parameters $\vec{\alpha}\in\mathbb{R}^N$. For low dimensions such as $N<10$ standard minimization routines such as Quasi-Newton optimization are recommended. For high dimension optimization problems stochastic gradient descent or simulated annealing are certainly preferable approaches. As starting values of these iterative optimization routines the ordinary least squares solutions can be utilized, if applicable.

Fifth, due to the choice of $f_{\vec{\varepsilon}_{l,r}}$ and $f_{\vec{\eta}_{h,r}}$ (in the best case representing the true data errors), the optimization problem can be more or less difficult. For example, selecting these densities with too small standard deviations, the objective function might contain a large number of non-distinct local extrema next to each other, which is difficult for local optimization algorithms. On the other side, selecting these densities with too large standard deviations may lead to very broad extrema, which can be helpful for the optimization algorithm but strongly reduces the information value of the observed data.

\subsubsection{Evaluation of Unpaired Data Subgroups}
\label{sec:eval_unpaired_subgroups}

Presenting the mathematical argumentation framework does not mean, that a practical model fitting problem at hand is well stated. Let us focus on the input set $D=\{\vec{x}_h, \;h=1,..,H\}$. We are choosing $R$ unpaired subgroups $S_{r}$ ($r=1,..,R$) which in total represents a partition of these input values
\begin{align*}
\bigcup\limits_{r=1}^{R} S_{r} = D\quad\wedge\quad S_{r}\cap S_{m} = \emptyset\;\;\forall\,r,m\in\{1,..,R\}, r\neq m\;.
\end{align*}
The question arises which partitioning is beneficial for the fit and which is not. At this point, we only want to discuss this problem by exploring the extremes:
\begin{itemize}
\item[A)] If all subgroups $S_r$ (approximately) contain a representative sample of the whole data set $D$, then the model fitting is qualitatively the same as if we would use the completely unpaired case, which can be regarded as useless for a practical model fit, since no useful pairing information is contained in such a partitioning.
\item[B)] If all subgroups $S_r$ are presenting different, separated areas of the input data set, i.e. each subgroup is very dissimilar to $D$.
\end{itemize}
In consequence, one way to judge about the practical usefulness of the partitioning is to look for dissimilarity of each $S_r$ to $D$ and between subgroups $S_r$. The question arises: What is a good measure to determine the \textit{dissimilarity} between $S_{r}$ and $D$ and between subgroups $S_r$ for all $r=1,..,R$? Only for high dissimilarity, the pairing inside the subgroups will not degrade the model fitting result strongly. In \textit{Supplement \ref{sec:res:plane_fit}} an illustrative example for this effect is presented.

For designing data observation processes with deliberately partially unpaired data (maybe due to privacy protection, or observational costs etc.) it could be helpful to measure such dissimilarities directly and we regard this as future work.

\subsubsection{Simultaneous Estimation of the Underlying Density Functions}
\label{sec:Est_Densities}

In Equation (\ref{equ:partiallyunordered_xy}) we assumed known and fixed input and output density functions $f_{\vec{\eta}_{h,r}}$ and $f_{\vec{\varepsilon}_{l,r}}$ for data groups $r=1,..,R$ with their element indices $h=1,..,H_r$ and $l=1,..,L_r$. An extension of the proposed estimation concept is that the density functions are depending on \textit{unknown parameters}, i.e. $\vec{\beta}_{h,r}$ for $\vec{x}$- and $\vec{\gamma}_{l,r}$ for $\vec{y}$-values, which we want to estimate simultaneously with the model parameters $\vec{\alpha}$. We denote this by density functions $f_{\vec{\eta}_{h,r}}(\;\cdot\; ; \vec{\beta}_{h,r})$ and $f_{\vec{\varepsilon}_{l,r}}(\;\cdot\;;\vec{\gamma}_{l,r})$. This leads to the extended Maximum Likelihood problem:
\begin{align*}
\text{argmax}_{\;(\vec{\alpha},\vec{\beta},\vec{\gamma})}\; \prod\limits_{r=1}^R \; \left[\;\frac{1}{L_r H_r}\sum\limits_{l=1}^{L_r} \sum\limits_{h=1}^{H_r} \;\int\limits_{\mathbb{R}^k} f_{\vec{\varepsilon}_{l,r}}( \vec{y}_{l,r}-\vec{M}(\vec{s};\vec{\alpha}) ; \vec{\gamma}_{l,r}) \cdot f_{\vec{\eta}_{h,r}}( \vec{x}_{h,r}-\vec{s} ; \vec{\beta}_{h,r}) \;\text{d}\vec{s}\,\right]\;.
\end{align*}

\textit{Remark:} One challenge of this extension is that the optimization gets high-dimensional with possibly many local maxima, which might occur due to the flexible interplay of densities with large standard deviations and the matching of the unpaired data groups. Further, too many parameters scaling with the data size may lead to problems of identifiability. In consequence, we assume that there will be the need to combine parameters and force the densities to smaller standard deviations, e.g. by penalizing parameters $(\vec{\beta},\vec{\gamma})$ which correspond to large standard deviations. There are many standard approaches for additive penalizing and we regard this out of the scope of the current presentation. A pragmatic approach for combining parameters could be to assume that for each input and output coordinate the error densities are known and identical with zero mean and unknown standard deviations, i.e. $\vec{\beta}:=\vec{\sigma}_{\eta}\in\mathbb{R}^k$ and $\vec{\gamma}:=\vec{\sigma}_{\varepsilon}\in\mathbb{R}^m$ (independent of $h,l$ and $r$). This allows a global estimation of the coordinate errors simultaneously to the model parameters and increases the degrees of freedom of the optimization problem only by $k+m$. For example, this corresponds for the \textit{errors-in-variables multiple linear regression} in Section \ref{sec:ex:EIV_MultLinReg_PartUnord} to additionally maximize all $\beta_n:=\sigma_{\eta,n}$ ($n=1,..,k$) and $\gamma:=\sigma_{\varepsilon}$, increasing the degrees of freedom from $k+1$ to $2\cdot(k+1)$.

\subsubsection{Bayesian Extension}

In this derivation, we defined the likelihood function, which we need to maximize in the previous sections by
\begin{align*}
\mathcal{L}(\vec{0}\,|\,\vec{\alpha}) := \prod\limits_{r=1}^R \;f_{ \vec{M}(\vec{X}^{\ast}_r;\vec{\alpha}) - \vec{Y}^{\ast}_r}(\vec{0})\;.
\end{align*}
The unusual perspective in this likelihood derivation is that our \textit{observation} is $\vec{0}$ since we must find the parameters $\vec{\alpha}$ of the difference random variable $\vec{M}(\vec{X}^{\ast}_r;\vec{\alpha}) - \vec{Y}^{\ast}_r$ to make $\vec{0}$ most likely. This can be extended to a classical \textit{Bayesian} perspective by introducing a prior for the random variable $\vec{\alpha} \sim \pi(\vec{\alpha})$. In consequence, the posterior density (utilizing Bayes' rule) gets
\begin{align*}
\pi(\vec{\alpha}\,|\,\vec{0}) &= c\cdot \mathcal{L}(\vec{0}\,|\,\vec{\alpha})\cdot\pi(\vec{\alpha})\\
&= c\cdot \left( \prod\limits_{r=1}^R \; \int\limits_{\mathbb{R}^k} f_{\vec{Y}^{\ast}_r }(\vec{M}(\vec{s};\vec{\alpha}))\cdot f_{\vec{X}^{\ast}_r}(\vec{s})\;\text{d}\vec{s}\right) \cdot\pi(\vec{\alpha}) \;,
\end{align*}
with $c$ a normalization constant. By utilizing a non- or weakly informative prior $\pi(\vec{\alpha})$ (such as $\pi(\vec{\alpha})=$ \textit{const.} on a large enough domain) the previously presented optimization problem is identical to the maximization of the posterior, leading to a \textit{Maximum A Posteriori} (MAP) estimate. This allows to interpret the plotted objective functions in the results Section \ref{sec:res_line} as presentations of the density functions of $\vec{\alpha}$, \textit{containing directly} the inherent estimation uncertainties about $\vec{\alpha}$ graphically as intensity maps. \medskip

\section{Simulation Study}

The purpose of the results section is to illustrate the presented general fitting approach by examples to improve understanding of the derived formulas.

\subsection{Demonstration for a Line Fit}
\label{sec:res_line}

\begin{figure}[h!]
\centering
\includegraphics[width=13cm]{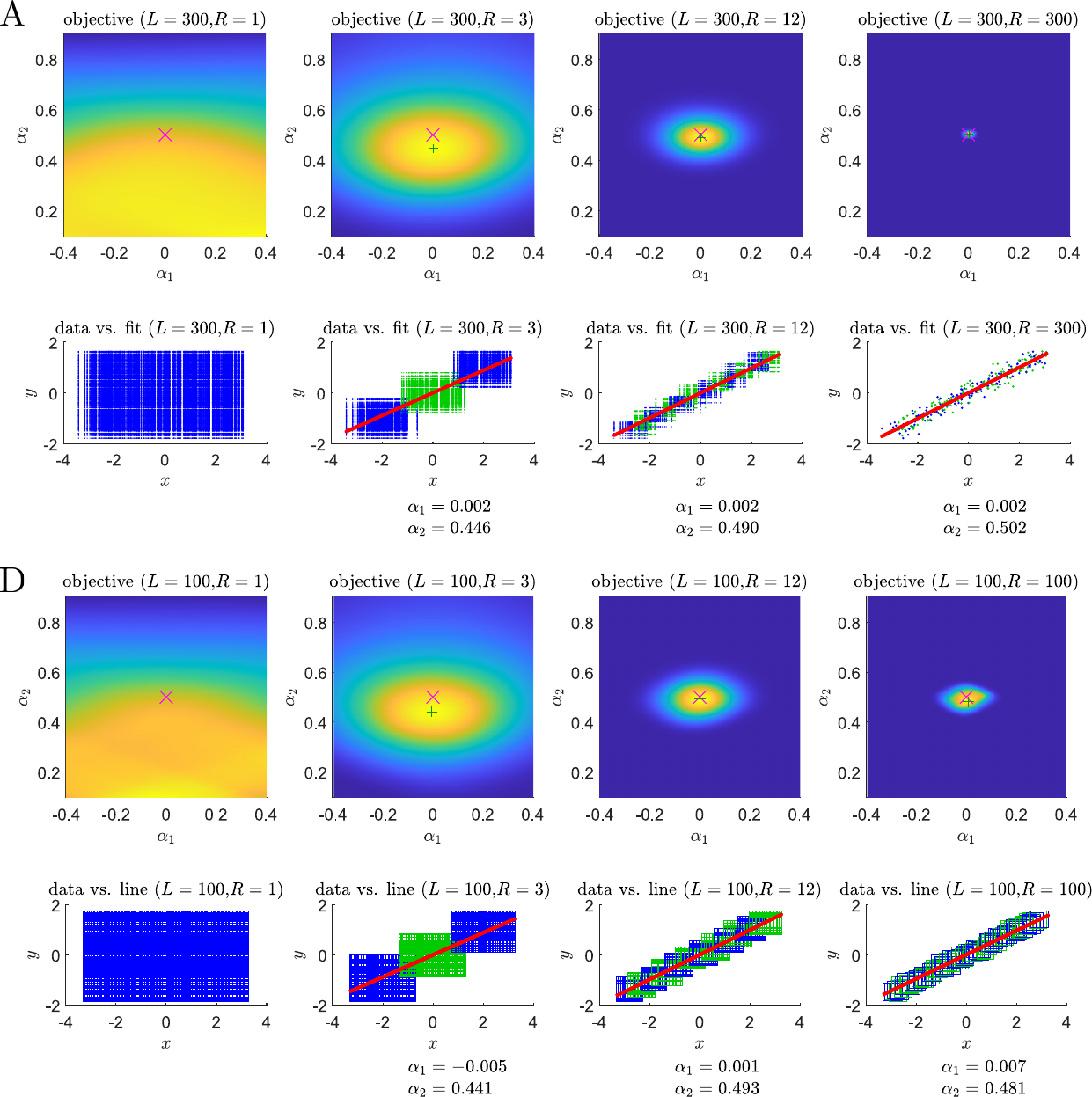}
\caption{Illustration of example results for different line fit scenarios according to scenarios $A$ (Gaussian error) and $D$ (interval data). For each case: top row: objective functions with true parameters (red crosses) and maximum (green crosses), bottom row: data presentation and line fit results. Inside the color-coded unpaired subgroups (green and blue) all possible correspondences are plotted. Columns: Four different scenarios of partial pairing with $R=1$, $R=3$, $R=12$ and $R=L=H$ groups.}  
\label{fig:Res:Gauss_Linear_centered}
\end{figure}

\begin{figure}[h!]
\centering
\includegraphics[width=13cm]{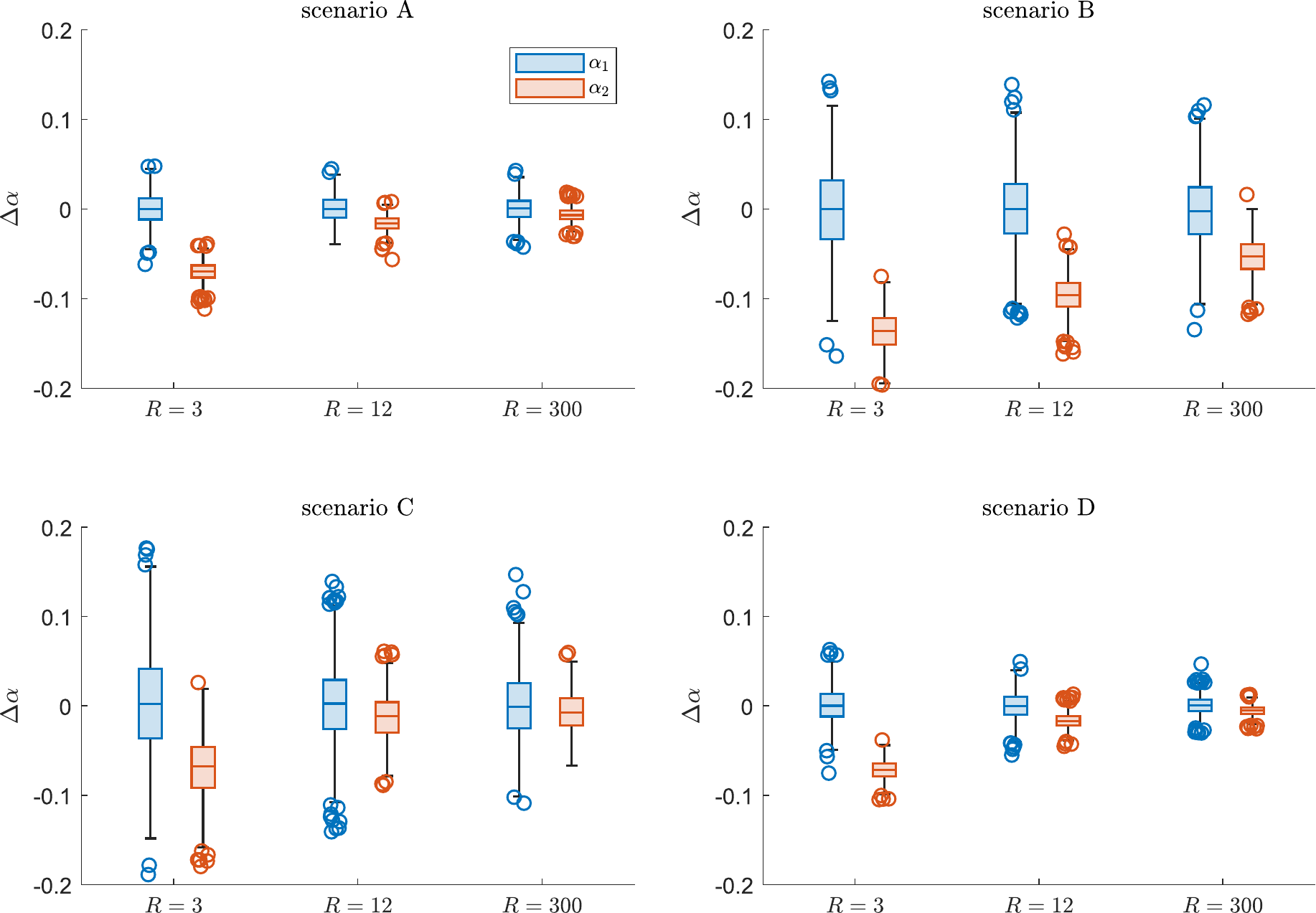}
\caption{Evaluation of scenarios $A$ to $D$ (with $L=300$) for $1000$ fits. Presented are the box plots of the residual errors $\Delta \alpha = \alpha_{\text{fit}} - \alpha_{\text{truth}}$ of the intercept $\alpha_1$ and slope $\alpha_2$. For each scenario different pairings are presented with $R=3$, $R=12$ and $R=L=H=300$.}  
\label{fig:Res:IntervalData_Linear}
\end{figure}

At first, a simple line fit is illustrated. For each of the following scenarios we vary the number of subgroups $R$. The following scenarios are investigated:
\begin{itemize}
\item \textit{Base scenario $A$}: The correct parameter values are $\alpha_1=0$ and $\alpha_2=0.5$. We utilize $L=H=300$ data points and the Gaussian data point disturbances are drawn with $\sigma_{\eta}=\sigma_{\varepsilon}=0.2$ and expectation $0$.  
\item \textit{Scenario $B$}: same as A, but with increased Gaussian disturbances $\sigma_{\eta}=\sigma_{\varepsilon}=0.6$.
\item \textit{Scenario $C$}: same as A, but with $L=H=36$.
\item \textit{Scenario $D$}: $L=H=100$ data points in Figure \ref{fig:Res:Gauss_Linear_centered} and $L=H=300$ in Figure \ref{fig:Res:IntervalData_Linear}. Interval data regression with uniform disturbances with standard deviations $\sigma_{\eta}=\sigma_{\varepsilon}=0.2$ and their corresponding interval boxes.
\end{itemize}

See Figure \ref{fig:Res:Gauss_Linear_centered} for presentation of example objective function and resulting line fits of scenario $A$ and $D$. 
The columns correspond to different numbers of unpaired subgroups $R\in\{1,3,12,L\}$. For the \textit{completeley unpaired} case ($R=1$) only the objective function is presented. 

See Figure \ref{fig:Res:IntervalData_Linear} for a systematic evaluation of the line fits for scenarios $A$ to $D$ utilizing $1000$ simulated fits with random data errors. Presented are the box plots of the residual errors for $\alpha_1$ and $\alpha_2$. 
Qualitative conclusions from Figures \ref{fig:Res:Gauss_Linear_centered} and \ref{fig:Res:IntervalData_Linear}: 
First, utilizing completely unpaired data leads to arbitrary insufficient results which can be observed by the non-distinct maxima of the objective functions. 
Second, the fewer subgroups $R$ are utilized, the broader (and more uncertain) gets the maximum in the objective function in Figure \ref{fig:Res:Gauss_Linear_centered}. Further, for very few subgroups, such as $R=3$, a bias on the slope $\alpha_2$ is introduced for all scenarios as presented in Figure \ref{fig:Res:IntervalData_Linear}. On the other side, it is obvious that reasonable estimation of the parameters is absolutely possible even if only partially unpaired data is available (comparing $R=12$ and $R=300$). 
Third, comparing scenarios $A$ to $B$: The uncertainty of estimation increases with an increased noise level of the data for all cases of $R$. 
Fourth, comparing scenarios $B$ to $C$: The higher noise level in the data leads to similar uncertainties in comparison to fewer data.  
Fifth, comparing scenarios $A$ to $D$: The main difference utilizing interval data compared to Gaussian disturbances is that also for the completely paired case no distinct maximum appears but a \textit{plateau} of high intensity values are observable in the objective function in Figure \ref{fig:Res:Gauss_Linear_centered}. In Figure \ref{fig:Res:IntervalData_Linear} it can be seen that interval data leads to similar results with a stronger bias on the slope for $R=3$.

In \textit{Supplement \ref{sec:res:plane_fit}} a plane fit is presented as an example for 2D input variables.

\subsection{Nonlinear Model with Anisotropic Observation Errors}
\label{sec:res_nonlinear}

\begin{figure}[h!]
\centering
\includegraphics[width=13cm]{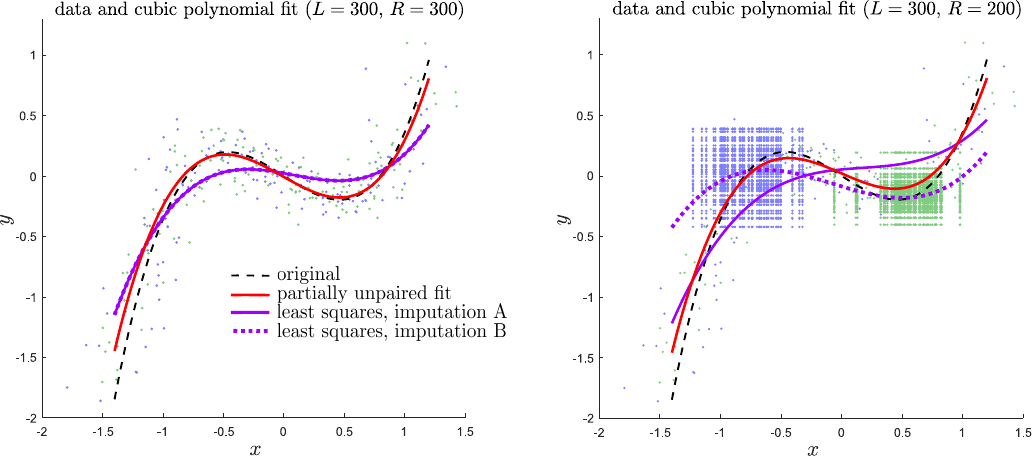}
\caption{Illustration of example results for cubic model fitting scenarios on data with Gaussian disturbances. The generating cubic function (black dashed line), the partially unpaired data model fitting (red line) and two simple comparison model fits (violet continuous and dotted lines) are presented. Left: completely paired case ($R=L=H$). Right: partially unpaired case with two unpaired areas with $51$ lost data correspondences each ($R=200$). Inside the color-coded unpaired groups (green or blue dots) all possible correspondences are plotted as dots.}  
\label{fig:Res:CubicData}
\end{figure}

In order to demonstrate the flexibility of this framework, the fitting of a nonlinear model $\mathbb{R} \mapsto \mathbb{R}$ ($k=1,m=1,N=4$) 
\begin{align*}
M(x;\vec{\alpha})=\alpha_1+\alpha_2\cdot x + \alpha_3\cdot x^2 + \alpha_4\cdot x^3\;, 
\end{align*}
is presented with anisotropic Gaussian disturbances $\sigma_{\eta}=0.2$ and $\sigma_{\varepsilon}=0.1$ for $L=H=300$ data points. In this model, nonlinearity holds with respect to $x$ and not with respect to $\vec{\alpha}$, which is deliberate and not necessary. In Figure \ref{fig:Res:CubicData} the results are presented for a completely paired case ($R=L=H$) (left) and a partially unpaired case (right) with two areas of lost pairing information. The general algorithm was implement according to section \ref{sec:numericals} with the simple trapezoidal rule for integration and the \textit{Nelder-Mead}-optimization in \textit{Matlab}.
For comparison of fitting results, Figure \ref{fig:Res:CubicData} shows the generating cubic function (black dashed line), the model fitting result according to Equation (\ref{equ:partiallyunordered_xy}) (red line)  and two simple comparison model fits (violet continuous and dotted lines). This comparison model fit is the ordinary least squares fit application of the cubic model (neglecting the noise in $x-$direction) with two different simple but intuitive \textit{imputation} treatments of the unpaired data: A) (= violet continuous line) Taking the average of the $x-$ and $y-$values as a new artificial data point in these two areas and else neglect the unpaired data. B) (= violet dotted line) Including all possible combinations of the unpaired data directly in the least squares fit. Both comparison approaches are regarded as suboptimal, but intuitive data \textit{imputations} for an unexperienced practitioner utilizing \textit{ordinary least squares}, and therefore, presented for demonstration.

In Figure \ref{fig:Res:CubicData} (left) the benefit of including an error model in $x$ additionally to $y$ is presented (\textit{ordinary least squares} does not contain an error model in $x$), showing clearly superior results of the fit (red line) compared to the overlapping violet lines. In Figure \ref{fig:Res:CubicData} (right) the performance of the both simple comparison model fits (violet lines) decrease significantly compared to the case on the left while the model fit (red line) stays robust, dealing in a stable way with the lost pairing information.

\section{Real Data Study: Life Expectancy}
\label{sec:RealDataStudy}

\begin{figure}[h!]
\centering
\includegraphics[width=13cm]{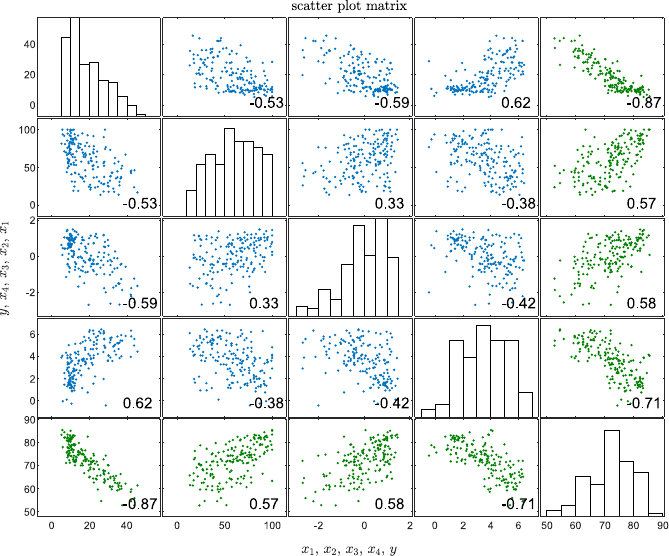}
\caption{Pair plot of the input data $x_1$ = \textit{Birth rate, crude (per 1,000 people)}, $x_2$ = \textit{Urban population (percent of total population)}, $x_3$ = \textit{Political Stability and Absence of Violence/Terrorism: Estimate}, $x_4$ = \textit{logarithm of Incidence of tuberculosis (per 100,000 people)} (blue plots) and the output variable $y$ = \textit{Life expectancy at birth, total (years)} (green plots). The Pearson correlation coefficient is presented in each correlation plot. On the diagonal are the histograms of each variable.}  
\label{fig:Res:PlotMatrix}
\end{figure}

It will be demonstrated how this framework can be utilized for a \textit{errors-in-variables multiple linear regression} problem with observational errors in $\vec{x}$ and $y$ on real data. This means, the model utilized is
\begin{align*}
M(\vec{x};\vec{\alpha})=\alpha_1+\sum\limits_{n=1}^{k} \alpha_{n+1}\cdot x_n\;.
\end{align*}
The task will be to fit this model to life expectancy data for most countries  in the world. The specific data set is taken from the \textit{world bank databank}\footnote{Data taken from \texttt{https://databank.worldbank.org/source/world-development-indicators} (June 2024), Database: World Development Indicators, data year 2020. } and utilizes $k=4$ input variables $x_1$ = \textit{Birth rate, crude (per 1,000 people) [SP.DYN.CBRT.IN]}, $x_2$ = \textit{Urban population (percent of total population) [SP.URB.TOTL.IN.ZS]}, $x_3$ = \textit{Political Stability and Absence of Violence/Terrorism: Estimate [PV.EST]}, $x_4$ = \textit{logarithm of Incidence of tuberculosis (per 100,000 people) [SH.TBS.INCD]} and the output variable $y$ = \textit{Life expectancy at birth, total (years) [SP.DYN.LE00.IN]}. The corresponding plot matrix is presented in Figure \ref{fig:Res:PlotMatrix} for all $192$ countries for which the variables were available.

As we want to demonstrate the \textit{errors-in-variables} approach, we need to define error densities for each variable. Since in the world bank data there are no error margins provided, we assume normally distributed errors with mean zero and standard deviations of $15\%$ of the standard deviation of the full data set for each variable $x_1\; (1.49),x_2\; (3.52),x_3\; (0.14),x_4\; (0.24)$ and $y\; (1.12)$.

For evaluation purposes, we perform a train-test-split with $172$ training countries and $20$ test countries in order to judge if we can learn from the training countries the life expectancy for the test countries based on the input variables. A major challenge is that the goodness of fit cannot be measured with the classical $R^2$ since it is only valid for cases with no errors in the input variables. In consequence, we utilize the extension of $R^2$ to the \textit{errors-in-variables} approach $R_{\delta}^2$ \citep{cheng_coefficient_2014} for multiple linear regression (see \textit{Supplement \ref{sec:app:R_delta_squared}} for more details).

\begin{figure}[h!]
\centering
\includegraphics[width=13cm]{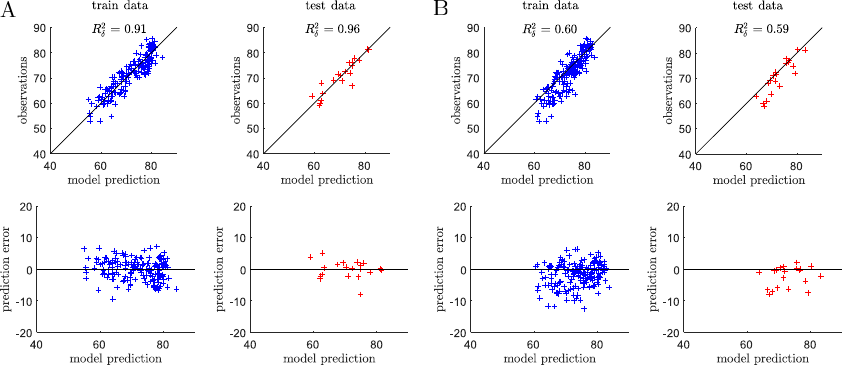}
\caption{Plot results for real and predicted values of the model fit. A) Utilizing all $172$ countries (completely paired) and B) utilizing only $R=11$ groups of countries (partially unpaired) with the $R_{\delta}^2$ values as goodness of fit measure. Top: Correlation plots, Bottom: Residual error plots. Left: Train data, Right: Test data.}  
\label{fig:Res:RD_Result001}
\end{figure}

First, we consider the case of the model fit with perfect pairing ($R=L=H$). Since the proposed algorithm was implemented with Monte Carlo methods as described in Section \ref{sec:numericals}, it was verified with an explicit solution for the multiple linear regression model \citep{cheng_coefficient_2014} and showed equivalent results. The results are presented in Figure \ref{fig:Res:RD_Result001}A as correlation and error plots of predicted and real output values for the train and test data sets. They show that the training allows a high quality prediction for the test countries with the multiple linear regression model for life expectancy. 
It is noted that the scatter plots themselves are defective, since the prediction is performed by taking the input values $\vec{x}$ assuming no errors. In order to illustrate also the errors in the input values, other plot types would need to be established. 

Second, one major approach in this work is to investigate explanatory power of the model fit if the train data is partially unpaired, i.e. if we only consider data of groups of countries rather than individual countries for model training. To achieve this, we introduce the country grouping along an additional criteria which is not part of the input: \textit{GDP per capita (current USD) [NY.GDP.PCAP.CD]} with increasing GDP per capita for each group. Please note, although the grouping is performed by GDP, the GDP itself is not part of the predictors and the overall information level for the predictors is reduced by this grouping compared to the completely paired data set. We considered $3$ different groupings with group sizes $L_r=H_r$ (the last group always consists of residual countries) of $4$ ($R=44$), $8$ ($R=22$) and $16$ ($R=11$). The results of the model fit based on this country groups are presented in Table \ref{tab:RealDataResults}. 
\begin{table}[h!]
\centering
\begin{tabular}{l|l|l|l}
number of groups & group sizes & $R_{\delta}^2$ (train) & $R_{\delta}^2$ (test) \\ \hline\hline
$172$ & $1$ & $0.91$ & $0.96$ \\ \hline
$44$ & $4$ & $0.74$ & $0.74$ \\ \hline
$22$ & $8$ & $0.70$ & $0.68$ \\ \hline
$11$ & $16$ & $0.60$ & $0.59$ 
\end{tabular}
\caption{Coefficient of determination $R_{\delta}^2$ in train and test data for different group sizes.}
\label{tab:RealDataResults}
\end{table}
The $R_{\delta}^2$ for train and test data is presented which show overall very high goodness-of-fit values but with decreasing values for increasing group sizes, as one might expect. Please note, the $R_{\delta}^2$ for testing contains a rather strong noise component since it is calculated by only $20$ data points in the test set. In addition, in Figure \ref{fig:Res:RD_Result001}B (for group size $16$) it is demonstrated that the fit results still align well for training and testing even for $11$ subgroups. It is noted, that the grouping is part of the training to get $\vec{\alpha}$, and not part of these scatter plots. This indicates that the model fit and the prediction for the test countries work quite well based only on the country groups, confirming that the pairing information can be reduced and this still provides a valuable model fit.

\section{Discussion and Conclusion}

In this work, we presented a general framework for model fitting scenarios with stochastic uncertainties for completely paired and partially unpaired data utilizing mixture models. The main advantage of this approach is its generality allowing for full flexibility about i) the number and dimensions of the data points, ii) the (possibly) individual error characteristics of each data point in the \textit{errors-in-variables} framework, iii) the type of (linear or nonlinear) model to be fitted, iv)  the specific level of pairing information, and v) completely avoiding \textit{ad-hoc} loss functions.  The presented loss function is derived from the data's pairing structure and data specific error characteristics, making it the most suitable fit for these problems. We present our framework as a generalization of \textit{total least squares} \citep{markovsky_overview_2007}, extending it to a broader \textit{errors-in-variables} context. By employing Gaussian errors with a line model, classical results are reproduced, but our approach also accommodates other scenarios, such as interval data through uniform distributions. The primary random variables in our study are the uncertainty variables $\vec{\eta}$ and $\vec{\varepsilon}$, with all stochastic derivations stemming directly from their definitions. This rigorous foundation is a key advantage of our approach.

The results in simulations and the real data study indicate by examples that there can be a trade-off between the level of pairing information (number and shape of unpaired subgroups) and estimation accuracy, leading to a problem specific practical saturation in the accuracy level one can achieve by utilizing partially paired data. This means, the information about the full pairing of data is not as important for the fitting process as one might think, and consequently, accurate results can be performed also with reduced pairing information. 
Reduced pairing information can be useful, for example, in cases where the data is partially corrupt, or by deliberately leaving out pairing information due to data privacy policies (e.g. anonymizing data by building unpaired subgroups).

We advocate to broaden the meaning of \textit{semi-supervised} learning, as we did in this paper for \textit{partially unpaired data}, in order to capture different scenarios of loss of pairing information which are practically relevant.

Although the presented framework might be general, the derived formulas lead only for specific selections of density function and model types to closed form solutions. The practical implementation can still be challenging, especially in cases with a high number of parameters $\vec{\alpha}$, leading to a high dimensional optimization problem of an objective function with possibly non-distinct or non-unique extrema. Further, if the involved probability densities are not leading to expressions where the integral in Equation (\ref{equ:partiallyunordered_xy}) can be exactly solved, numerical approximations of these integrals can be challenging expecially for high-dimensional input data.

In the paper, it is only briefly presented how the ML approach can be directly extended applying MAP approaches. This was done to directly interpret the plotted likelihood functions in the results section as posterior densities by utilizing a non- or weakly informative prior. With this interpretation, we are able to directly quantify the uncertainties of the parameters $\vec{\alpha}$ of the model fit, allowing the calculation of credibility intervals or regions.

In general, Maximum Likelihood estimators are asymptotically consistent under some conditions, like identifiabilty. This extends to our framework, for fully paired data. However, for partially unpaired data, one would first need a meaningful definition of how the data and the data subgroups grow towards infinity. This is strongly related to the question of how the \textit{dissimilarity} of unpaired subgroups of the data can be measured, as discussed in Section \ref{sec:eval_unpaired_subgroups}. This remains an interesting open question with certainly a differentiated answer which we direct to future work on this topic.

Although the presentation of model fitting in this paper had regression problems in mind, the same argumentation can be applied to classification tasks. The adaption is that the output data $\vec{y}$ and the image of $\vec{M}(\cdot;\vec{\alpha})$ is discrete and finite. Further discretization, such as discrete density functions $\vec{f}_{\vec{\varepsilon}_l}$ can be modeled by Dirac distributions in order to directly apply the presented equations, e.g. applying the \textit{sifting property} for the obtained integral in Equation (\ref{equ:partiallyunordered_xy}).

In total, this is a general argumentation framework for model fitting with many possible applications and an introduction to the specific treatment for partially unpaired data. The focus of this presentation is on the applied researcher, explaining all derivations and results in detail as well as providing numerical implementation strategies and interpretations of numerical examples. Further work is encouraged in order to extend this framework or provide further examples (e.g., benchmarking compared to alternative fitting methods) of expressive applications.

\appendix
\section{Completely Paired Data: Derivations for Errors in $\vec{y}$ only}
\label{sec:Deriv_errors_in_y_only}

In this case, the stochastic disturbances are only present in the output data: 
\begin{align}
\vec{y}_l^{\,\ast} := \vec{y}_l - \vec{\varepsilon}_l
\end{align}
with $\vec{y}_l^{\,\ast}\in\mathbb{R}^m$ the random variable of the true value, and the uncertainty random variable $\vec{\varepsilon}_l \sim f_{\vec{\varepsilon}_l}(\vec{s}) : \mathbb{R}^m\mapsto \mathbb{R}$ independent for all $l=1,..,L$. In a Bayesian context, the observed and true input values are the same $\vec{x}_l = \vec{x}_l^{\,\ast}$.
With this, we introduce the technical argumentation of model fitting by
\begin{align}
\vec{M}(\vec{x}_l;\vec{\alpha}) &= \vec{y}_l^{\,\ast} \quad\forall\,l=1,..,L\\
\vec{M}(\vec{x}_l;\vec{\alpha}) &= \vec{y}_l - \vec{\varepsilon}_l \quad\forall\,l=1,..,L\;,
\end{align}
i.e. for given (undisturbed) $\vec{x}_l$ we want to predict the true value $\vec{y}_l^{\,\ast}$. The first step in this technical presentation is to bring all basic random variables to the left side and equal this to $\vec{0}$:
\begin{align}
\vec{M}(\vec{x}_l;\vec{\alpha}) - \vec{y}_l + \vec{\varepsilon}_l &= \vec{0} \quad\forall\,l=1,..,L\;.
\end{align}
We follow the interpretation: due to $\vec{\varepsilon}_l$ being a random variable, the left side is interpreted as a shifted random variable which density function value should have highest value at $\vec{0}\in\mathbb{R}^m$, following the idea of Maximum Likelihood for the parameters $\vec{\alpha}$.
\begin{align}
\Rightarrow \quad&\text{argmax}_{\;\vec{\alpha}}\quad f_{ \bigcap\limits_{l=1}^L \left[\, \vec{M}(\vec{x}_l;\vec{\alpha}) - \vec{y}_l + \vec{\varepsilon}_l \,\right]}(\vec{0})\\
\quad{=}\quad& \text{argmax}_{\;\vec{\alpha}}\quad \prod\limits_{l=1}^L f_{\vec{M}(\vec{x}_l;\vec{\alpha}) - \vec{y}_l + \vec{\varepsilon}_l }(\vec{0})\qquad\text{\small (independence of $\vec{\varepsilon}_l$)}\\
\quad{=}\quad& \text{argmax}_{\;\vec{\alpha}}\quad \prod\limits_{l=1}^L f_{\vec{\varepsilon}_l}(\vec{y}_l-\vec{M}(\vec{x}_l;\vec{\alpha}))\;.\label{equ:ordered_y}\qquad\text{\small (shifted $\vec{\varepsilon}_l$)}
\end{align}
We recognize this as the common standard result of Maximum Likelihood (ML) in this new way of technical argumentation and we present standard examples in the following.

\subsubsection*{Example: Gaussian Disturbance}
\label{sec:OnlyY_Gaussian}

Introducing Gaussian disturbances, we get $\vec{\varepsilon}_l \sim \mathcal{N}(\vec{0},\,\sigma_{\vec{\varepsilon}}^{2}\cdot I_{m\times m})(\vec{s})\;\;\forall\, l=1,..,L$ this results in
\begin{align}
\quad{\Rightarrow}\quad& \text{argmax}_{\;\vec{\alpha}}\quad \prod\limits_{l=1}^L \text{e}^{-\frac{1}{2\,\sigma_{\vec{\varepsilon}}^2} ||\vec{y}_l-\vec{M}(\vec{x}_l;\vec{\alpha})||^2}\qquad\text{\small (inserting pdf)}\\
{=}\quad& \text{argmax}_{\;\vec{\alpha}}\quad  \text{e}^{-\frac{1}{2\,\sigma_{\vec{\varepsilon}}^2} \sum\limits_{l=1}^L ||\vec{y}_l-\vec{M}(\vec{x}_l;\vec{\alpha})||^2}\\
{=}\quad& \text{argmin}_{\;\vec{\alpha}}\quad  \sum\limits_{l=1}^L ||\vec{y}_l-\vec{M}(\vec{x}_l;\vec{\alpha})||^2
\end{align}
which is the case of multivariate (nonlinear) ordinary least squares.

\subsubsection*{Example: Fitting a Line and Gaussian Disturbance (Linear Regression)}

Further utilizing the one-dimensional affine model $M(x;\vec{\alpha})=\alpha_1+\alpha_2\cdot x$ and inserting it, we get
\begin{align}
\Rightarrow \quad& \text{argmin}_{\;\vec{\alpha}}\quad  \sum\limits_{l=1}^L (y_l-\alpha_1-\alpha_2\cdot x_l)^2
\end{align}
which has the classical unique solution of the normal equations of ordinary least squares leading to a fitted line with parameters

\begin{align}
\alpha_1 = \frac{ \overline{x^2}\cdot\overline{y} - \overline{x}\cdot\overline{xy}}{\overline{x^2}-\overline{x}^2},\quad \alpha_2 = \frac{\overline{xy} - \overline{x}\cdot\overline{y}}{\overline{x^2}-\overline{x}^2}
\end{align}
with
\begin{align}
\overline{x} := \frac{1}{L}\sum\limits_{l=1}^L x_l &,\quad \overline{y} := \frac{1}{L}\sum\limits_{l=1}^L y_l\\
\overline{x^2} := \frac{1}{L}\sum\limits_{l=1}^L x^2_l &,\quad \overline{xy} := \frac{1}{L}\sum\limits_{l=1}^L x_l\cdot y_l\;.
\end{align}

\section{Completely Paired Data: Relation to Deming Regression}
\label{sec:Examples_errors_in_x_and_y}

Deming regression is equivalent to the maximum likelihood for independent normally distributed observation errors in $x_l$ and $y_l$ ($l=1,..,L$), i.e. ${\eta}_l \sim \mathcal{N}(0,\,\sigma_{{\eta}}^{2})(s)\;\forall l=1,..,L$ and ${\varepsilon}_l \sim \mathcal{N}(0,\,\sigma_{{\varepsilon}}^{2})(s)\;\forall l=1,..,L$ with a line model. Since the true value $s_l$ for $x_l$ is unknown just like the parameters $\vec{\alpha}$ they are estimated by maximizing them simultaneously with the parameters $\vec{\alpha}$, which leads to the effect that the estimation of the true values influences the estimation of the parameters $\vec{\alpha}$: 
\begin{align*}
\text{argmax}_{\;\vec{\alpha}, s_1,..,s_L}\quad \prod\limits_{l=1}^L \;\text{e}^{-\frac{1}{2\,\sigma_{{\varepsilon}}^2}(y_l-\alpha_1-\alpha_2\cdot s_l)^2 -\frac{1}{2\,\sigma_{{\eta}}^2} (x_l-s_l)^2 }
\end{align*}
In the case of this paper, we are also estimating the best parameters $\vec{\alpha}$ but independently of any specific true value $s_l$: We are averaging over all possible true values by the use of the law of total probability, compare Equation (\ref{equ:deming_alternative_01}), which can be interpreted as an \textit{integrated Deming regression}. This is a valid alternative perspective leading to a slightly more defensive estimation of $\vec{\alpha}$ which is not influenced by the estimation of $s_l$. An interesting observation is that the second part of derived objective function in Equation (\ref{equ:deming_alternative})
\begin{align}
\text{argmin}_{\;\vec{\alpha}}\quad \sum\limits_{l=1}^L\; \frac{(\alpha_1+\alpha_2\cdot x_l - y_l)^2}{2\,(\alpha_2^2\,\sigma_{{\eta}}^2 + \sigma_{{\varepsilon}}^2)}\label{equ:deming_derivation}\;,
\end{align}
actually leads to the \textit{classical Deming equations} and the first part of Equation (\ref{equ:deming_alternative}) $\frac{L}{2}\,\ln\left( \alpha_2^2\,\sigma_{{\eta}}^2 + \sigma_{{\varepsilon}}^2\right)$ can be interpreted as a \textit{penalty} added to the classical Deming regression, showing the more defensive estimation approach. Especially for large $\sigma_{\eta}$ the parameter $\alpha_2$ will tend slightly more to zero. The derivation of the classical Deming regression in this context is directly the minimization of Equation (\ref{equ:deming_derivation}) by setting the gradient to zero
\begin{align}
\nabla_{\vec{\alpha}} \sum\limits_{l=1}^L\; \frac{(\alpha_1+\alpha_2\cdot x_l - y_l)^2}{2\,(\alpha_2^2\,\sigma_{{\eta}}^2 + \sigma_{{\varepsilon}}^2)} = \vec{0}\;,
\end{align}
whose solution results in the classical Deming regression coefficients
\begin{align}
\alpha_1 &= \frac{1}{L}\sum\limits_{l=1}^L\;y_l-\alpha_2\cdot x_l\\
\alpha_2 &= \frac{s_{yy} - \frac{\sigma_{\varepsilon}^2}{\sigma_{\eta}^2}\cdot s_{xx} + \sqrt{ \left(s_{yy} - \frac{\sigma_{\varepsilon}^2}{\sigma_{\eta}^2}\cdot s_{xx}\right)^2 + 4\,\frac{\sigma_{\varepsilon}^2}{\sigma_{\eta}^2}\,s_{xy}^2 } }{2\,s_{xy}}
\end{align}
with 
\begin{align}
s_{xx} = \frac{1}{L}\,\sum\limits_{l=1}^L (x_l - \overline{x})^2,\quad s_{yy} = \frac{1}{L}\,\sum\limits_{l=1}^L (y_l - \overline{y})^2,\quad s_{xy} = \frac{1}{L}\,\sum\limits_{l=1}^L (x_l - \overline{x})\cdot (y_l - \overline{y})\;.
\end{align}

In conclusion, the presented approach leads for \textit{Deming type problems} to formulas, which we call \textit{integrated Deming regression}. These formulas can be interpreted as penalized \textit{classical Deming regression} showing the more defensive approach by averaging over the true value during estimation of $\vec{\alpha}$ compared to estimating them simultaneously in \textit{classical Deming regression}.

\section{Demonstration for a Plane Fit with Gaussian Disturbance for Partially Unpaired Data}
\label{sec:res:plane_fit}

One possible part of demonstrating the flexibility of this framework is to show how it works in higher dimensions, which will be indicated by the previously introduced plane fit model. We are using $L=H=1600$ data points $\vec{x}_h\in\mathbb{R}^2$ and $y_l\in\mathbb{R}$ and numbers of subgroups are $R\in\{6,18,100,L=H=1600\}$. The data generation parameters are the same as for the base scenario $A$ in the line fit section but with the true values $\alpha_1=0$, $\alpha_2=0.2$ and $\alpha_3=0.4$. 

Obviously the partitioning of the data has much more possibilities due to a much richer neighboring information for $\vec{x}_h$ in 2D. In consequence, the discussed dissimilarity of the groups to the total data set gets more difficult to study. Nonetheless, it can be demonstrated that the utilization with more pairing information does increase accuracy and meaningful estimation can be performed even with a low number of groups.

\begin{figure}[h!]
\centering
\includegraphics[width=13cm]{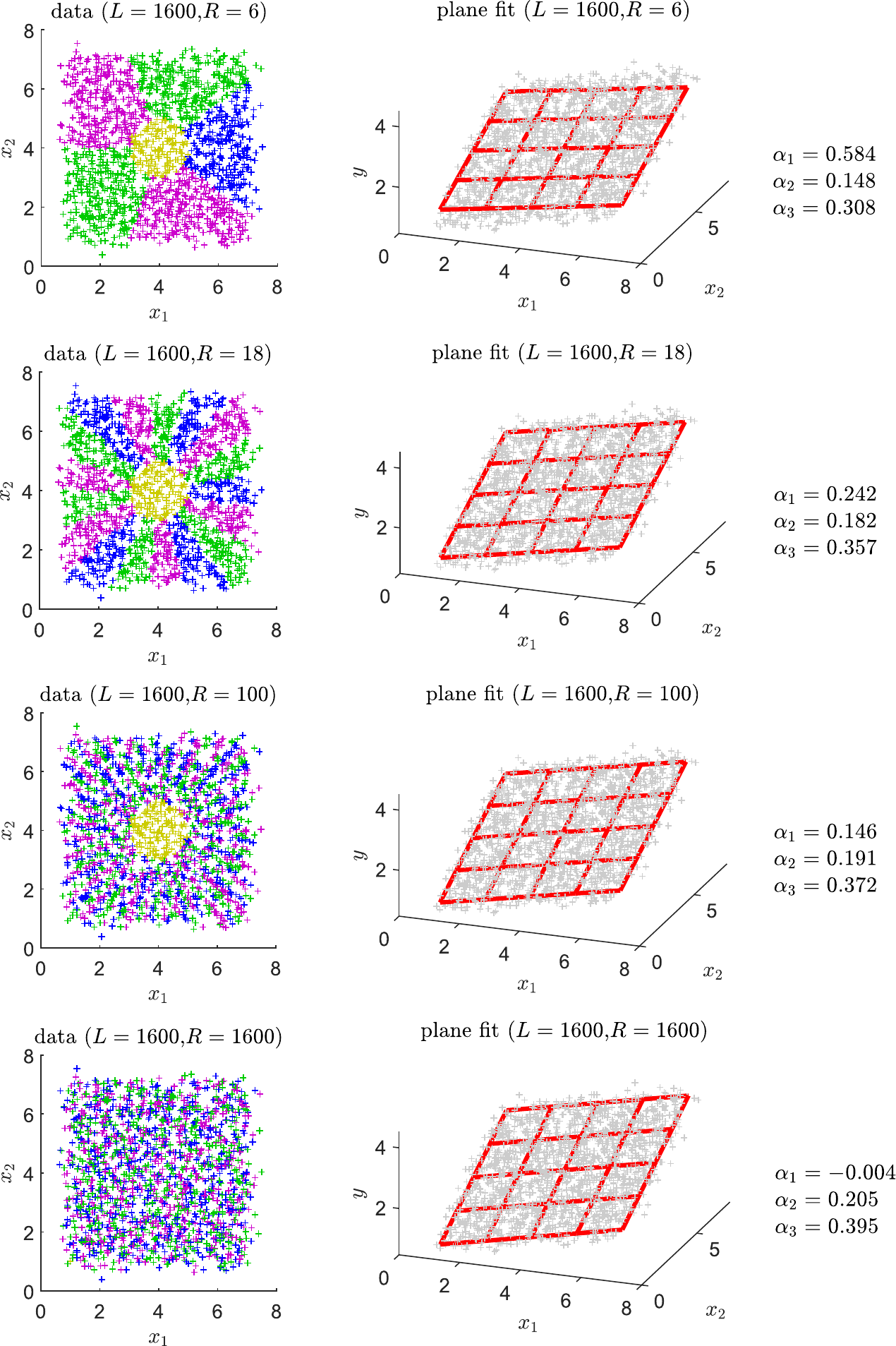}
\caption{Illustration of example results for different plane fitting scenarios utilizing a partitioning of separated groups. Four different scenarios of partial pairing with $R=6$, $R=18$, $R=100$ and $R=L=H$ unpaired subgroups. For each case: Left plot: representing the unpaired data subgroups by color-coding. Right plot: presenting the fitted plane (red) in the full data (gray).}  
\label{fig:Res:Gauss_Linear_Plane}
\end{figure}

\begin{figure}[h!]
\centering
\includegraphics[width=12.5cm]{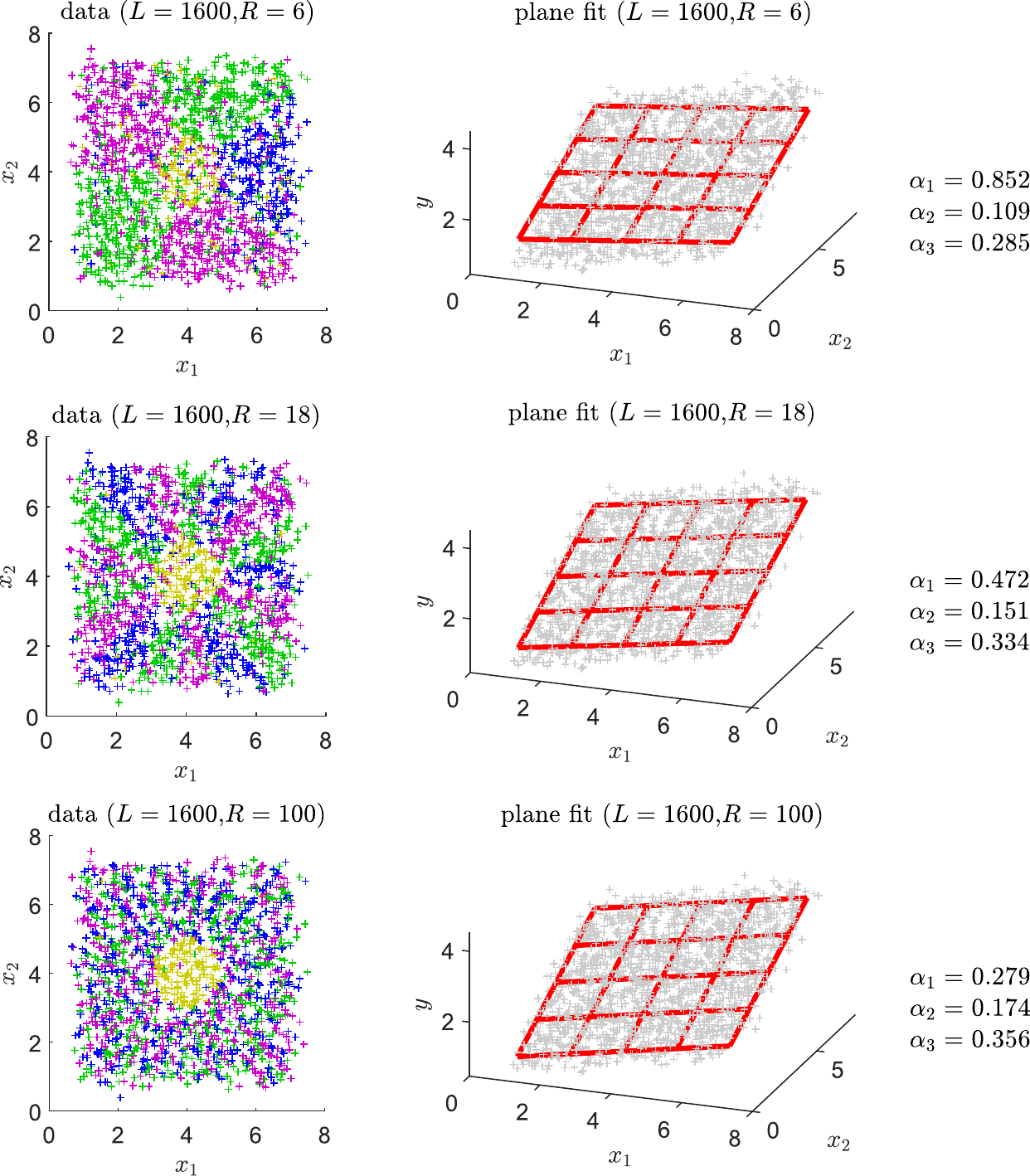}
\caption{Illustration of example results for different plane fitting scenarios utilizing a partitioning of overlapping groups (i.e., randomly switching the group label for approximately $30\%$ of the data points compared to Figure \ref{fig:Res:Gauss_Linear_Plane}). Three different scenarios of partial pairing with $R=6$, $R=18$ and $R=100$ unpaired subgroups are demonstrated in the same type of presentation as in Figure \ref{fig:Res:Gauss_Linear_Plane}.}  
\label{fig:Res:Gauss_Linear_Plane2}
\end{figure}

In Figure \ref{fig:Res:Gauss_Linear_Plane} the fitting results are presented for separated groups (no overlapping of data groups in the $\vec{x}$-plane). In addition, in Figure \ref{fig:Res:Gauss_Linear_Plane2} the same data is utilized but a random relabeling/switching of group labels is performed for approximately $30\%$ of the data, which makes all groups slightly more similar to each other. As proposed in \textit{Section \ref{sec:eval_unpaired_subgroups}}, the plane fitting results get worse the more similar the grouping gets.


\section{Application of $R_{\delta}^2$}
\label{sec:app:R_delta_squared}

In \citep{cheng_coefficient_2014} a consistent goodness of fit measure for \textit{errors-in-variables} multiple linear regression is presented if the standard deviations of the input variables are known. We show its direct application in this appendix. With $X$ the $L\times k$-Matrix of $L$ observations and $k$ predictor variables, $\vec{y}$ the $L\times 1$ vector of output observations, $\Sigma_{\delta}$ the $k\times k$ covariance matrix of predictor variables, $S=\frac{1}{L}\,X^T\, P\, X$ and $P=I_{L\times L}-\frac{1}{L} \mathbf{1}_{L\times L}$ ($\mathbf{1}$ being the matrix consisting of $1$s) the goodness of fit is defined by
\begin{align*}
R_{\delta}^2 = \text{min}\left( \;\frac{\vec{b}^T\, S \, \vec{b} }{\frac{1}{L}\,\vec{y}^T\, P\, \vec{y} + \vec{b}^T\, \Sigma_{\delta}\, \vec{b} }  \;,\;1\;\right)\;,
\end{align*}
where $\vec{b}$ is the vector of fitted slopes, i.e., in the notation of the multiple linear regression model of this paper $\vec{b} = (\alpha_2,\alpha_3,\dots,\alpha_k)^T$ being independent of the intercept $\alpha_1$.

% Bibliography
\bibliography{stochmodfitmm}

\begin{thebibliography}{20}
\providecommand{\natexlab}[1]{#1}
\providecommand{\url}[1]{\texttt{#1}}
\expandafter\ifx\csname urlstyle\endcsname\relax
  \providecommand{\doi}[1]{doi: #1}\else
  \providecommand{\doi}{doi: \begingroup \urlstyle{rm}\Url}\fi

\bibitem[Zhang(1997)]{zhang_parameter_1997}
Zhengyou Zhang.
\newblock Parameter estimation techniques: a tutorial with application to conic
  fitting.
\newblock \emph{Image and Vision Computing}, 15\penalty0 (1):\penalty0 59--76,
  January 1997.
\newblock \doi{10.1016/S0262-8856(96)01112-2}.

\bibitem[Bishop(2006)]{bishop_pattern_2006}
Christopher~M. Bishop.
\newblock \emph{Pattern recognition and machine learning}.
\newblock Information science and statistics. Springer, New York, 2006.
\newblock ISBN 978-0-387-31073-2.

\bibitem[Wang et~al.(2022{\natexlab{a}})Wang, Ma, Zhao, and
  Tian]{wang_comprehensive_2022}
Qi~Wang, Yue Ma, Kun Zhao, and Yingjie Tian.
\newblock A {Comprehensive} {Survey} of {Loss} {Functions} in {Machine}
  {Learning}.
\newblock \emph{Annals of Data Science}, 9\penalty0 (2):\penalty0 187--212,
  April 2022{\natexlab{a}}.
\newblock \doi{10.1007/s40745-020-00253-5}.

\bibitem[Hoegele et~al.(2013)Hoegele, Loeschel, Dobler, Koelbl, and
  Zygmanski]{hoegele_bayesian_2013}
Wolfgang Hoegele, Rainer Loeschel, Barbara Dobler, Oliver Koelbl, and Piotr
  Zygmanski.
\newblock Bayesian {Estimation} {Applied} to {Stochastic} {Localization} with
  {Constraints} due to {Interfaces} and {Boundaries}.
\newblock \emph{Mathematical Problems in Engineering}, 2013:\penalty0 1--17,
  2013.
\newblock \doi{10.1155/2013/960421}.

\bibitem[Bai and Hsing(2005)]{bai_broken_2005}
Zhidong Bai and Tailen Hsing.
\newblock The broken sample problem.
\newblock \emph{Probability Theory and Related Fields}, 131\penalty0
  (4):\penalty0 528--552, April 2005.
\newblock \doi{10.1007/s00440-004-0384-5}.

\bibitem[Liang et~al.(2007)Liang, Mukherjee, and West]{liang_use_2007}
Feng Liang, Sayan Mukherjee, and Mike West.
\newblock The {Use} of {Unlabeled} {Data} in {Predictive} {Modeling}.
\newblock \emph{Statistical Science}, 22\penalty0 (2), May 2007.
\newblock \doi{10.1214/088342307000000032}.

\bibitem[Wang et~al.(2022{\natexlab{b}})Wang, Tang, and Ye]{wang_paired_2022}
Yudong Wang, Yanlin Tang, and Zhi-Sheng Ye.
\newblock Paired or {Partially} {Paired} {Two}-sample {Tests} {With}
  {Unordered} {Samples}.
\newblock \emph{Journal of the Royal Statistical Society Series B: Statistical
  Methodology}, 84\penalty0 (4):\penalty0 1503--1525, September
  2022{\natexlab{b}}.
\newblock \doi{10.1111/rssb.12541}.

\bibitem[Kostopoulos et~al.(2018)Kostopoulos, Karlos, Kotsiantis, and
  Ragos]{kostopoulos_semi-supervised_2018}
Georgios Kostopoulos, Stamatis Karlos, Sotiris Kotsiantis, and Omiros Ragos.
\newblock Semi-supervised regression: {A} recent review.
\newblock \emph{Journal of Intelligent \& Fuzzy Systems}, 35\penalty0
  (2):\penalty0 1483--1500, August 2018.
\newblock \doi{10.3233/JIFS-169689}.

\bibitem[Qi and Luo(2022)]{qi_small_2022}
Guo-Jun Qi and Jiebo Luo.
\newblock Small {Data} {Challenges} in {Big} {Data} {Era}: {A} {Survey} of
  {Recent} {Progress} on {Unsupervised} and {Semi}-{Supervised} {Methods}.
\newblock \emph{IEEE Transactions on Pattern Analysis and Machine
  Intelligence}, 44\penalty0 (4):\penalty0 2168--2187, April 2022.
\newblock \doi{10.1109/TPAMI.2020.3031898}.

\bibitem[Bennett(2001)]{bennett_how_2001}
Derrick~A. Bennett.
\newblock How can {I} deal with missing data in my study?
\newblock \emph{Australian and New Zealand Journal of Public Health},
  25\penalty0 (5):\penalty0 464--469, October 2001.
\newblock \doi{10.1111/j.1467-842X.2001.tb00294.x}.

\bibitem[Sterne et~al.(2009)Sterne, White, Carlin, Spratt, Royston, Kenward,
  Wood, and Carpenter]{sterne_multiple_2009}
J.~A~C Sterne, I.~R White, J.~B Carlin, M.~Spratt, P.~Royston, M.~G Kenward,
  A.~M Wood, and J.~R Carpenter.
\newblock Multiple imputation for missing data in epidemiological and clinical
  research: potential and pitfalls.
\newblock \emph{BMJ}, 338\penalty0 (jun29 1):\penalty0 b2393--b2393, September
  2009.
\newblock \doi{10.1136/bmj.b2393}.

\bibitem[Michael et~al.(2020)Michael, Miljkovic, and
  Melnykov]{michael_mixture_2020}
Semhar Michael, Tatjana Miljkovic, and Volodymyr Melnykov.
\newblock Mixture modeling of data with multiple partial right-censoring
  levels.
\newblock \emph{Advances in Data Analysis and Classification}, 14\penalty0
  (2):\penalty0 355--378, June 2020.
\newblock \doi{10.1007/s11634-020-00391-x}.

\bibitem[McCaw et~al.(2022)McCaw, Aschard, and Julienne]{mccaw_fitting_2022}
Zachary~R. McCaw, Hugues Aschard, and Hanna Julienne.
\newblock Fitting {Gaussian} mixture models on incomplete data.
\newblock \emph{BMC Bioinformatics}, 23\penalty0 (1):\penalty0 208, December
  2022.
\newblock \doi{10.1186/s12859-022-04740-9}.

\bibitem[Hoegele(2024{\natexlab{a}})]{hoegele_stochastic-geometrical_2024-1}
Wolfgang Hoegele.
\newblock A {Stochastic}-{Geometrical} {Framework} for {Object} {Pose}
  {Estimation} {Based} on {Mixture} {Models} {Avoiding} the {Correspondence}
  {Problem}.
\newblock \emph{Journal of Mathematical Imaging and Vision}, June
  2024{\natexlab{a}}.
\newblock \doi{10.1007/s10851-024-01200-2}.

\bibitem[Hoegele(2024{\natexlab{b}})]{hoegele_investigating_2024}
Wolfgang Hoegele.
\newblock Combinatorial potential of random equations with mixture models:
  Modeling and simulation, March 2024{\natexlab{b}}.
\newblock arXiv:2403.20152 [cs, math, stat].

\bibitem[Deming(1964)]{deming_statistical_1964}
W.~Edwards Deming.
\newblock \emph{Statistical adjustment of data}.
\newblock Dover publ, New York, unabridged and corr. republication edition,
  1964.
\newblock ISBN 978-0-486-64685-5.

\bibitem[Markovsky and Van~Huffel(2007)]{markovsky_overview_2007}
Ivan Markovsky and Sabine Van~Huffel.
\newblock Overview of total least-squares methods.
\newblock \emph{Signal Processing}, 87\penalty0 (10):\penalty0 2283--2302,
  October 2007.
\newblock \doi{10.1016/j.sigpro.2007.04.004}.

\bibitem[Lima~Neto and De~Carvalho(2008)]{lima_neto_centre_2008}
Eufrásio De~A. Lima~Neto and Francisco De~A.T. De~Carvalho.
\newblock Centre and {Range} method for fitting a linear regression model to
  symbolic interval data.
\newblock \emph{Computational Statistics \& Data Analysis}, 52\penalty0
  (3):\penalty0 1500--1515, January 2008.
\newblock \doi{10.1016/j.csda.2007.04.014}.

\bibitem[Souza et~al.(2017)Souza, Souza, Amaral, and
  Silva~Filho]{souza_parametrized_2017}
Leandro~C. Souza, Renata~M.C.R. Souza, Getúlio~J.A. Amaral, and Telmo~M.
  Silva~Filho.
\newblock A parametrized approach for linear regression of interval data.
\newblock \emph{Knowledge-Based Systems}, 131:\penalty0 149--159, September
  2017.
\newblock \doi{10.1016/j.knosys.2017.06.012}.

\bibitem[Cheng et~al.(2014)Cheng, {Shalabh}, and Garg]{cheng_coefficient_2014}
C.-L. Cheng, {Shalabh}, and G.~Garg.
\newblock Coefficient of determination for multiple measurement error models.
\newblock \emph{Journal of Multivariate Analysis}, 126:\penalty0 137--152,
  April 2014.
\newblock \doi{10.1016/j.jmva.2014.01.006}.

\end{thebibliography}

\end{document}